\documentclass[11pt,a4paper]{article}

\usepackage{jheppub}
\usepackage{mathrsfs,ulem}

\def\S{{\mathbb S}}

\newcommand{\alg}[1]{\mathfrak{#1}}

\newcommand{\el}{\nonumber\\}

\def\ads{{\rm AdS}_5\times {\rm S}^5}

\def\xp{{x_1^+}}
\def\xm{{x_1^-}}
\def\xpp{{x_2^+}}
\def\xmm{{x_2^-}}

\def\DE{{\Delta E}}
\def\DF{{\Delta F}}
\def\DH{{\Delta H}}

\def\DHF{{\Delta \hat F}}

\def\tU{{\widetilde U}}
\def\tV{{\widetilde V}}

\newcommand{\tl}[1]{\tilde{#1}}
\newcommand{\wtl}[1]{\widetilde{#1}}

\def\DOE{{\Delta^{\!op} E}}
\def\DOF{{\Delta^{\!op} F}}

\def\dM{{\delta M}}

\newcommand{\stateA}[1]{|#1\rangle^{\rm{I}}}
\newcommand{\stateB}[1]{|#1\rangle^{\rm{II}}}
\newcommand{\stateC}[1]{|#1\rangle^{\rm{III}}}
\newcommand{\costateA}[1]{ ^{\rm{I}}\langle #1}
\newcommand{\costateB}[1]{ ^{\rm{II}}\langle #1}

\newcommand{\ket}[1]{|#1\rangle}

\def\a{\mathsf{a}}
\def\ad{\mathsf{a}^\dag}

\def\afQ{\widehat{\cal{Q}}}

\title{The Bound State S-matrix of the Deformed Hubbard Chain}

\author[a]{Marius de Leeuw}
\author[b,c]{Takuya Matsumoto}
\author[d,e]{Vidas Regelskis}

\affiliation[a]{Max-Planck-Institut f\"ur 
Gravitationsphysik\\Albert-Einstein-Institut\\Am M\"uhlenberg 1, 14476 Potsdam,
Germany}
\affiliation[b]{School of Mathematics and Statistics, University of Sydney, \\
NSW 2006, Australia}
\affiliation[c]{Graduate School of Mathematics, Nagoya University,\\Nagoya
464-8602, Japan}
\affiliation[d]{Department of Mathematics, University of York,\\Heslington, York
YO10 5DD, UK}
\affiliation[e]{Institute of Theoretical Physics and Astronomy of Vilnius
University,\\Go\v{s}tauto 12, Vilnius 01108, Lithuania}

\emailAdd{mdeleeuw@aei.mpg.de}
\emailAdd{m05044c@math.nagoya-u.ac.jp}
\emailAdd{vr509@york.ac.uk}

\abstract{In this work we use the $q$-oscillator formalism to construct the
atypical (short) supersymmetric representations of the centrally extended
$\mathcal{U}_q(\alg{su}(2|2))$ algebra. We then determine the S-matrix
describing the scattering of arbitrary bound states. The crucial ingredient in
this derivation is the affine extension of the aforementioned algebra.}

\begin{document}

\begin{flushright}\small{AEI-2011-062}\end{flushright}

\maketitle


\section{Introduction}

Integrable systems constitute a special class of models in mathematics and
physics. Their properties allow them to be solved exactly and thus they
form a very useful playground for studying various systems. One common
feature shared by these models is that they are closely related to some
underlying algebraic structures. Hence most of the quantum integrable
systems exhibit a large and powerful symmetry algebra, for example
of Yangian or quantum affine type. A particularly interesting example is
the Hubbard model. 

The Hubbard model, which was named after John Hubbard, is the simplest model of
interacting particles on a lattice. It has only two terms in the Hamiltonian:
the hopping term (kinetic energy) and the Coulomb potential \cite{Hubbard}. The
model describes an ensemble of particles in a periodic potential at sufficiently
low temperatures such that all the particles may be considered to be in the
lowest Bloch band. Moreover, any long-range interactions between the
particles are considered to be weak enough and are consequently ignored.
It is based on the tight-binding approximation of superconducting systems and
the motion of electrons between the atoms of a crystalline solid. Despite its
apparent simplicity, there are different applications and generalizations
describing a plethora of interesting phenomena. In the case when interactions
between particles on different sites of the lattice can not be neglected and are
taken into account, the model is often referred to as the Extended
Hubbard model. The particles can either be fermions, as in Hubbard's original
work, or bosons, and the model is then referred as either the
Bose-Hubbard model or the boson Hubbard model. The latter can be
used to study systems such as bosonic atoms on an optical lattice (for a decent
overview of various generalizations see reprint volumes \cite{Hub1, Hub2, Hub3}
and also a more recent book \cite{Hub4}).

A very specific class of models that share features with the
one-dimensional Hubbard model and the supersymmetric t-J model \cite{tJ} is the
so-called Alcaraz and Bariev model \cite{AB}. It contains an extra spin-spin
interaction term in the Hamiltonian and it shows some characteristics of
superconductivity. This model can be viewed as a quantum deformation of the
Hubbard model in much the same way as the Heisenberg XXZ model is a quantum
deformation of the XXX model. This model has a specific R-matrix which can not
be written as a function of the difference of two associated spectral
parameters. This paradigm is related to the very interesting but at the same
time complicated algebraic properties of the model. 

In recent years there has been renewed interest in integrable models arising
from the discovery of integrable structures in the context of the AdS/CFT
correspondence. For a recent review see \cite{AdSReview} and references therein.
The worldsheet S-matrix encountered there is one of the central objects of
research and it turns out to have a lot in common with the specific cases of the
Hubbard model considered in \cite{BAnalytic,BK}. Interestingly, the S-matrix of
such a Hubbard model is obtained as a special limit of this worldsheet S-matrix
\cite{MartinsMelo}.

The exact integrability of the one-dimensional Hubbard model was established by
B.~Shastry \cite{Shastry}. It was also shown that the model exhibits
$\mathcal{Y}(\alg{su}(2))\oplus\mathcal{Y}(\alg{su}(2))$ Yangian symmetry
\cite{UK}. However this symmetry is insufficient to constrain Shastry's S-matrix
completely. Similarly, the worldsheet S-matrix for the $\ads$ superstring also
turns out to have Yangian symmetry \cite{BeisertYangian}. However the Yangian in
this model is based on a larger Lie algebra, the centrally extended
$\mathfrak{su}(2|2)$ Lie superalgebra. This underlying Lie superalgebra
turns out to be powerful enough to constrain the S-matrix
\cite{BeisertFundamental,AFZ,BJ} (up to an overall phase, the so-called
`dressing factor' \cite{Janik,AFS,BES}) in the case where at least one of the
representations is fundamental. However, Yangian symmetry (or equivalently the
Yang-Baxter equation) is required in order to find the S-matrix describing the
scattering of states that live in higher representations \cite{AFBound,MdLrmat}.
This specifically concerns the bound states in the system which transform
in supersymmetric short representations \cite{Bsu22,Dorey,CDO,CDO2}. The
bound state scattering matrix can be explicitly constructed with the help of the
underlying Yangian symmetry \cite{ALT}.

Nevertheless, there are still some problems concerning this infinite dimensional
Yangian algebra due to some of its unusual features. The centrally
extended $\mathfrak{su}(2|2)$ Lie superalgebra has a degenerate Cartan matrix
which prohibits the direct application of most of techniques related to the
theory of Yangians. For the case at hand this has been partially circumvented in
several ways: by enlarging the algebra by an $\alg{sl}(2)$ automorphisms
\cite{BeisertYangian}, by considering the $\alpha\to0$ limits of the exceptional
Lie superalgebra $\mathfrak{d}(2,1;\alpha)$ \cite{MM} or building
Drinfeld's second realization \cite{ST}. However this still proves to be an
obstacle when, for example, one tries to construct the universal R-matrix
\cite{ALTlong,ALTblocks}. This object encodes all the scattering data in the
theory in a purely algebraic form. Another issue that is not completely
understood is the appearance of the so-called secret symmetry \cite{MMT}. This
is an additional symmetry of the S-matrix that does not have a 
corresponding Lie algebraic generator in the extended $\alg{su}(2|2)$ 
and could be interpreted as a $\alg{u}(1)$ outer automorphism of
$\alg{su}(2|2)$. Resolving these issues could shed some light on the  underlying algebraic
structures and put the methods used to solve the model on a more firm
footing.

A possible route for attacking these issues was put forward in \cite{BK}, where 
the quantum deformation $\mathcal{Q}$ of the extended $\alg{su}(2|2)$ algebra 
was studied. This $q$-deformed algebra has a number of interesting features such
as a rather symmetric realization of the different central elements.
Excitingly, just as in the non-deformed case, there is a link to Hubbard
models, more specifically it describes spectrum of deformed supersymmetric
one-dimensional Hubbard models \cite{DFFR,BK}. The undeformed Hubbard
model is revealed by taking a specific limit of deformed model \cite{Bcl}.
Moreover, by sending the quantum deformation parameter $q\to1$, the S-matrix
under the consideration reduces to the AdS/CFT worldsheet S-matrix. As such, this matrix encompasses both
different varieties of Hubbard models as well as the AdS/CFT
worldsheet S-matrix and seems to provide a unifying algebraic framework
for describing this class of models.

The $q$-deformed S-matrix in the fundamental representation is constrained up to
an overall phase by requiring invariance under $\mathcal{Q}$ itself. However, in
the light that both the AdS/CFT and the Hubbard model S-matrices are actually invariant under an
infinite dimensional symmetry algebra, it should not be surprising that such a
structure is also present here. Indeed, the larger algebraic structure
underlying this S-matrix is the quantum affine algebra $\afQ$ \cite{BGM}. This
infinite dimensional algebra is obtained by adding an additional fermionic node
to the Dynkin diagram of $\mathcal{Q}$. In the $q\to1$ limit one can
retrieve the Yangian generators of centrally
extended $\alg{su}(2|2)$ by considering the appropriate combinations of
generators of $\afQ$. This fuels the idea that $\afQ$ plays a similar role as the Yangian in the
undeformed case. More specifically, it is expected that the S-matrix in the
higher representations is uniquely defined up to an overall phase by the
underlying quantum affine algebra $\afQ$. This indeed turns out to be the case
as we will show in this work.

The class of representations we are considering in this work are the
supersymmetric short representations. These representations are called short
because the central elements are not independent; they satisfy the so-called
shortening condition \cite{Bsu22}. In order to construct these representations,
we employ the formalism of quantum oscillators. It is a
quantum version of the well-known harmonic oscillator algebra and is defined as
\begin{align*}
&[N,\a]=-\a, && [N,\ad]=\ad, && \a\,\ad- q\,\ad \a = q^{-N}.
\end{align*}
The use of quantum oscillators in the context of quantum groups was investigated
earlier in \cite{Macfarlane,Biedenharn,Hayashi}. By
employing Fock space type modules, $q$-oscillators naturally give rise to the
representations of quantum groups. This approach was first formulated for the quantum deformed algebra
$\mathcal{U}_q(\alg{sl}(2))$ and later extended
to simple Lie (super)algebras of a more general type,
see e.g.\ \cite{Chaichian}. Since then quantum oscillators have become an
important part of the theory of quantum deformed algebras.

Apart from being an interesting mathematical playground for studying
the quantum affine algebra $\afQ$ and its S-matrix, there is also a more
elaborate motivation for considering these representations and the corresponding
S-matrix. Firstly, there might be some possible applications in the context of
the deformed Hubbard model. Secondly, it turns out that bound state states
transform exactly in these representations of $q$-oscillator algebra. It is
important to study bound states for many reasons. For example, bound
states usually play a crucial role in the thermodynamics of the model. In the
case of the non-deformed model in AdS/CFT, the thermodynamic Bethe ansatz (TBA)
formalism is key in describing the complete spectrum of the theory
\cite{AFstring,GKV,BFT,AFTBA}. The bound state S-matrix then governs the large
volume solutions of both the TBA equations and the Y-system. Thus this is one of
the first steps towards the TBA and Y-system formalism for the $q$-deformed
model. And, consequently, it might give some useful insights in these structures
in the context of the AdS/CFT superstring. For example, there might be an interesting link to
the recently constructed $q$-deformed Pohlmeyer reduced version of the
superstrings in the $\ads$ background \cite{HT,HHM} which seems to be
closely related to the $q$-deformed model constructed in \cite{BK}.

In this work we derive the matrix structure of the general bound state
S-matrix by employing the methods used in the context of the AdS/CFT 
superstring \cite{ALT}, but rather than using the Yangian
symmetry we make use of the underlying quantum affine algebra $\afQ$. 
Our approach is based on the identification of invariant subspaces in the
scattering theory that are specified by their invariance properties under the
Cartan elements of the algebra. Then we use the rest of the algebra generators
to relate these subspaces to each other resulting in the explicit
form of the corresponding S-matrix. Just as in \cite{ALT} we find the S-matrix
in a factorized form reminiscent of the Drinfeld twist \cite{Drinfeld}.  

The paper is organized as follows. 
In section 2 we discuss the quantum deformation $\mathcal{Q}$ of the extended
$\mathcal{U}(\alg{su}(2|2))$ algebra and its affine extension $\afQ$. 
Then in section 3 we introduce the quantum oscillator formalism and construct
the supersymmetric short representations of $\afQ$. 
In section 4 we present the explicit derivation of the S-matrix for these
representations. 
Subsequently, in section 5, we specify some explicit cases, we reproduce the
fundamental R-matrix and also we give the precise form of the scattering matrix
when one of the spaces forms a fundamental representation. We end with a brief
discussion on the results and interesting directions for future research. The
majority of the S-matrix coefficients and results of the intermediate steps of
the performed calculations are spelled out in the appendices. 


\section{Quantum affine algebra of extended \texorpdfstring{$\mathcal{U}_q(\alg{su}(2|2))$}{U(su22)}}\label{sec;AffineAlgebra}

In this section we review the quantum deformation of the extended
$\alg{su}(2|2)$ algebra \cite{BK} and its affine extension \cite{BGM}. 


\subsection{Quantum deformation of extended \texorpdfstring{$\alg{su}(2|2)$}{su22}}

The quantum deformed extended $\alg{su}(2|2)$ algebra $\mathcal{Q}$ was
introduced in \cite{BK}.  This algebra is generated by the three sets of
Chevalley-Serre generators $\{E_j,K_j,F_j\}$ ($j=1,2,3$) where $E_j$ and $F_j$
are raising and lowering generators respectively and $K_j=q^{H_j}$ are the
Cartan generators. We will consider the case when $E_2$ and $F_2$ are fermionic
generators and the rest are bosonic. This corresponds to the $\alg{su}(2|2)$
Dynkin diagram in Figure \ref{fig:OXO}. 
\begin{figure}\centering
\includegraphics{FigDynkinOXO.mps}
\caption{Dynkin diagram for the $\alg{su}(2|2)$ algebra.}
\label{fig:OXO}
\end{figure}
In addition, this algebra has two central charges $U$ and $V=q^C$ and two
parameters: the deformation parameter $q$ and the coupling constant $g$. There
is also a third parameter $\alpha$, which describes the relative scaling of $E_2$ and $F_2$. 
Even though it is possible absorb this parameter into thes generators by a suitable redefinition, 
we will keep it unspecified.  
 

\paragraph{Algebra.}

The commutation relations which include the mixed Chevalley-Serre generators are
($j,k=1,2,3$) 
\begin{align}
\label{nonaff:mix}
K_jE_k=q^{+DA_{jk}}E_kK_j, \qquad K_jF_k=q^{-DA_{jk}}F_kK_j ,\qquad 
[E_j,F_k\}=D_{jj}\delta_{jk}\frac{K_j-K^{-1}_j}{q-q^{-1}},  
\end{align}
where the associated Cartan matrix $A$ and normalization matrix $D$ are given by 
\begin{align}
DA=\left(
\begin{array}{rrr}
+2 & -1 & 0 \\
-1 & 0 & +1 \\
0 & +1 & -2 \\
\end{array}\right),
\qquad
D=\mathrm{diag}(+1,-1,-1)\;.
\end{align}
There are also the unmixed commutation relations, called the Serre relations ($j=1,3$),  
\begin{align}
\label{nonaff:serre}
&[E_1,E_3]=\{E_2,E_2\}=\bigl[E_j,[E_j,E_2]\bigr]-(q-2+q^{-1})E_jE_2E_j=0, 
\el
&[F_1,F_3]=\{F_2,F_2\}=\bigl[F_j,[F_j,F_2]\bigr]-(q-2+q^{-1})F_jF_2F_j=0. 
\end{align}
In addition, this algebra satisfies the extended Serre relations that give rise
to two central elements $U$ and $V$ as follows,  
\begin{align}
\label{nonaff:extserre}
g\alpha(1-U^2V^2) &=\bigl\{[E_2,E_1],[E_2,E_3]\bigr\}-(q-2+q^{-1})E_2E_1E_3E_2,
\nonumber\\ 
g\alpha^{-1}(V^{-2}-U^{-2})
&=\bigl\{[F_2,F_1],[F_2,F_3]\bigr\}-(q-2+q^{-1})F_2F_1F_3F_2. 
\end{align}
The central element $V$ is also related to the Cartan generators through   
\begin{align}
V^{-2}&=K_1K_2^2K_3
\;.
\end{align}
The conventional $\mathcal{U}_q(\alg{su}(2|2))$ algebra is obtained in the
limit $g\to0$.


\paragraph{Coalgebra.}

The defining relations of $\mathcal{Q}$ are compatible with the following
coalgebra structure. The coproduct of the group like elements $X\in\{U,V,K\}$ is
${\rm \Delta}(X)=X\otimes X$ and the coproducts of the Chevalley-Serre
generators $E_j$ and $F_j$ ($j=1,3$) take the standard forms. However the
coproducts of the fermionic generators $E_2$ and $F_2$ involve an additional
braiding factor $U$, which is one of the central charges of the algebra alluded
to in the previous paragraph,
\begin{align}
\label{nonaff:copro}
{\rm\Delta}(E_j)=E_j\otimes1 + K^{-1}_jU^{+\delta_{j,2}}\otimes E_j,
\qquad 
{\rm \Delta}(F_j)=F_j\otimes K_j + U^{-\delta_{j,2}}\otimes F_j. 
\end{align}
The coalgebra can be extended to a Hopf algebra. We will
give the relevant definitions of the antipode and counit later on.


\subsection{Affine Extension}

\begin{figure}\centering
\includegraphics{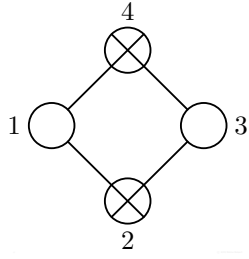}
\caption{Dynkin diagram for the affine $\widehat{\alg{su}}(2|2)$ algebra.}
\label{fig:OXOX}
\end{figure}

The infinite dimensional quantum affine algebra $\afQ$ is the affine extension
of ${\mathcal{Q}}$ introduced in \cite{BGM}. The affine extension is obtained by
adding an additional node into the Dynkin diagram as depicted in Figure
\ref{fig:OXOX}. 
The remarkable property of this diagram is that the additional fermionic node is
a copy of the second node. Therefore, we introduce the affine Chevalley-Serre
generators $\{E_4,F_4,K_4\}$ as copies of $\{E_2,F_2,K_2\}$ and assume that they satisfy the same
commutation relations as are given in \eqref{nonaff:mix}, \eqref{nonaff:serre}
and \eqref{nonaff:extserre} and also have the same coalgebra structure
\eqref{nonaff:copro}. 
Thus, we introduce an additional set of the parameters $g, \alpha$ and central
charges $U, V$. 
We distinguish these two sets by adhering subscripts to them arising from the
generators to which they are associated,
\begin{align}
g\to g_k, \qquad \alpha \to \alpha_k, \qquad U\to U_k, \qquad V\to V_k, 
\qquad \text{with}\qquad k=2,4.
\end{align}
Next, we need to determine the commutation relations $\{E_2,F_4\}$ and
$\{E_4,F_2\}$ in such way that they would be compatible with the coalgebra
structure,
\begin{align}
{\rm\Delta}(\{E_2,F_4\})=\{{\rm\Delta}(E_2),{\rm\Delta}(F_4)\}
\qquad\text{and}\qquad 
{\rm\Delta}(\{E_4,F_2\})=\{{\rm\Delta}(E_4),{\rm\Delta}(F_2)\}\;.
\end{align}
%


\paragraph{Algebra.}

As a result, we obtain the quantum affine algebra $\afQ$ \cite{BGM}. 
The mixed commutation relations of it are given by ($i,j=1,3$)
\begin{align}
\label{aff:mix}
K_iE_j &= q^{+DA_{ij}}E_jK_i, & K_iF_j &= q^{-DA_{ij}}F_jK_i, \nonumber\\
\{E_2,F_4\} &= -\tilde{g}\tilde{\alpha}^{-1}(K_4 - U_2 U_4^{-1}K_2^{-1}), &
\{E_4,F_2\} &=\tilde{g}\tilde{\alpha}(K_2 - U_4 U_2^{-1}K_4^{-1}), \nonumber\\
[E_j,F_j \} &= D_{jj} \frac{K_j-K_j^{-1}}{q-q^{-1}}& [E_i,F_j \} &= 0, 
\quad \text{for}~~i\neq j,\ i+j\neq 6\;. 
\end{align}
with the two new constants $\tilde g$ and $\tilde \alpha$ and the associated
supersymmetric Cartan matrix $A$ and normalization matrix $D$ given by 
\begin{align}
DA=\left( \begin{array}{rrrr}
+2 & -1 & 0 & -1\\
-1 & 0 & +1 & 0\\
0 & +1 & -2 & +1\\
-1 & 0 & +1 & 0
\end{array}\right),
\qquad
D = \mathrm{diag}(1,-1,-1,-1).
\end{align}
These are supplemented by the following Serre relations ($j=1,3$ and $k=2,4$)
\begin{align}
&[E_1,E_3] = E_2 E_2 = E_4 E_4 = \{E_2,E_4\} =0\;, \el
&[F_1,F_3] = F_2 F_2 = F_4 F_4 = \{F_2,F_4\} =0\;, \el
&[E_j,[E_j,E_k]]-(q-2+q^{-1})E_jE_kE_j = 0\;, \el 
&[F_j,[F_j,F_k]]-(q-2+q^{-1})F_jF_kF_j = 0 \;.
\end{align}
The central charges are related to the quartic Serre relations as ($k=2,4$) 
\begin{align}\label{eqn;quarticSerre}
g_k\alpha_k(1-U_k^2V_k^2)
&=\bigl\{[E_k,E_1],[E_k,E_3]\bigr\}-(q-2+q^{-1})E_kE_1E_3E_k\;, \nonumber\\ 
g_k\alpha_k^{-1}(V_k^{-2}-U_k^{-2})
&=\bigl\{[F_k,F_1],[F_k,F_3]\bigr\}-(q-2+q^{-1})F_kF_1F_3F_k\;.
\end{align}
and the central charges $V_k$ are related with Cartan charges through ($k=2,4$) 
\begin{align}
V^{-2}_k&=K_1K_k^2K_3\;. 
\end{align}
%


\paragraph{Coalgebra.}

The group-like elements $X \in \{1,K_j,U_k,V_k\}$ ($j=1,2,3,4$ and $k=2,4$) have
the coproduct ${\rm \Delta}$, the antipode ${\rm S}$ and the counit
$\varepsilon$ defined in the usual way,
\begin{equation}
\Delta(X)=X\otimes X, \qquad S(X)=X^{-1}, \qquad \varepsilon(X)=1, 
\end{equation}
while the coproducts of the Chevalley-Serre generators are deformed by the
central elements $U_k$ as follows ($j=1,2,3,4$),
\begin{align}
\Delta(E_j) &= E_j\otimes 1+K_j^{-1} U_{2}^{+\delta_{j,2}} U_{4}^{+\delta_{j,4}}
\otimes E_j, &
{\rm S}(E_j) &= -U_{2}^{-\delta_{j,2}} U_{4}^{-\delta_{j,4}} K_j E_j,  &
\varepsilon(E_j) &= 0, 
\nonumber\\
\Delta(F_j) &= F_j\otimes K_j+ U_{2}^{-\delta_{j,2}} U_{4}^{-\delta_{j,4}}
\otimes F_j, &
{\rm S}(F_j) &= -U_{2}^{+\delta_{j,2}} U_{4}^{+\delta_{j,4}} F_jK_j^{-1}, &
\varepsilon(F_j) &= 0.
\end{align}
It is important to note that the above coproducts are compatible with all the
defining relations, including the commutators $\{E_2,F_4\}$ and $\{E_4,F_2\}$ in
\eqref{aff:mix}. The opposite coproduct is defined as $\Delta^{\!op} = \mathcal
P\,\Delta\,\mathcal P$ with $\mathcal P$ being the graded permutation operator.


\paragraph{Parameter constraints.}

In general, the quantum affine algebra $\afQ$ has seven parameters $g_k,
\alpha_k, \tilde\alpha, \tilde g, q$ ($k=2,4$). 
A suitable choice of them which lead to an interesting fundamental
representation was performed in \cite{BGM}:
\begin{equation}
g_2=g_4=g, \qquad 
\alpha_2=\alpha_4\,\tilde\alpha^{-2}=\alpha,\qquad 
\tilde g^2=\frac{g^2}{1-g^2(q-q^{-1})^2}\,.  
\end{equation}
This choice of parameters is also compatible with the bound state
representations.
Thus in this paper we only consider the quantum affine algebra $\afQ$, 
parametrized by four independent parameters $g, \alpha, \tilde\alpha, q$ given
in the relations above. 


\section{Quantum oscillators and representations}

In this section we will provide all the necessary background for constructing
the bound state S-matrix for the $q$-deformed Hubbard model. We will build the
bound state representation by introducing $q$-oscillator formalism linking it to
the aforementioned quantum affine algebra.


\subsection{q-Oscillators}

We first introduce the notion of $q$-oscillators and discuss how to obtain the
representations of the quantum deformed algebras using $q$-oscillators. A
concise overview of the $q$-oscillators and their relation to such
representations may be found in \cite{ChaichianBook,Petersen}. 


\paragraph{Definitions.}

The $q$-oscillator ($q$-Heisenberg-Weyl algebra) $\mathcal{U}_q(\alg{h}_4)$ is
the associative unital algebra consisting of the generators
$\{\ad,\a,w,w^{-1}\}$ that satisfy the following relations,
\begin{align}\label{eqn;DefRelQosc}
w\,\ad &= q\,\ad w, && q w\,\a = \a\,w,\\
w w^{-1} &=  w^{-1} w =1, &&\a\,\ad - q\,\ad \a = w^{-1}.\nonumber
\end{align}
From the defining relations one can see that the element $w^{-1}(\ad \a
-\frac{w-w^{-1}}{q-q^{-1}})$ is central. As such, we will set it to zero in the
remainder. Then one easily obtains
\begin{align}\label{eqn;DefRelQosc2}
&\ad \a = \frac{w-w^{-1}}{q-q^{-1}}, &&\a\,\ad =
\frac{qw-q^{-1}w^{-1}}{q-q^{-1}}.
\end{align}
We will also need to consider the fermionic version of the $q$-oscillator. The
above notion is extended to include fermionic operators by adjusting the
defining relations in the following way (we keep the same notation for bosonic
and fermionic $\a,\,\ad$ for now)
\begin{align}
w\,\ad &= q\,\ad w, && q w\,\a  = \a\,w,\\
w w^{-1} &=  w^{-1} w =1, &&\a\,\ad + q\,\ad \a = w.\nonumber
\end{align}
In this case, the central element is $w(\ad \a -\frac{w-w^{-1}}{q-q^{-1}})$.
Again we set this element to zero, resulting in the following identities
\begin{align}\label{eqn;oscferm}
&\ad \a = \frac{w-w^{-1}}{q-q^{-1}}, &&\a\,\ad = \frac{qw^{-1} -
q^{-1}w}{q-q^{-1}}.
\end{align}
Of course in the fermionic case the operators $\a,\ad$ square to zero.
Equation \eqref{eqn;oscferm} implies that this only is consistent if $w^2 =
1,q^2$. Below we will identify $w\equiv q^N$, where $N=0,1$ is the number of
fermions making it indeed compatible.


\paragraph{Fock space.}

The $q$-oscillator algebra can be used to define representations of
$\mathcal{U}_q(\alg{sl}(2))$ in a very simple way. Let us first build the Fock
representation of $\mathcal{U}_q(\alg{h}_4)$. For this purpose consider a vacuum
state $|0\rangle$ such that 
\begin{align}
\a|0\rangle = 0, 
\end{align}
then the Fock vector space $\cal{F}$ generated by the states of the form
\begin{align}
|n\rangle = (\ad)^n|0\rangle\,,
\end{align}
is an irreducible module of $\mathcal{U}_q(\alg{h}_4)$. Let us first consider
the bosonic $q$-oscillators. With the help of the defining relations
\eqref{eqn;DefRelQosc} and \eqref{eqn;DefRelQosc2} one finds that the action of
the oscillator algebra generators on this module is
\begin{align}
&\ad|n\rangle = |n+1\rangle, && \a|n\rangle = [n]_q|n-1\rangle,&& w|n\rangle =
q^n|n\rangle.
\end{align}
This makes it natural to identify $w \equiv q^N$, where $N$ is understood as a
number operator. Analogously, fermionic generators are found to act as
\begin{align}
&\ad|n\rangle = |n+1\rangle, && \a|n\rangle = [2-n]_q|n-1\rangle,&& w|n\rangle =
q^n|n\rangle.
\end{align}
However, due to the fermionic nature, $n$ can only take the values $0$ and $1$
and thus the identity $[2-n]_q = [n]_q$ holds.

Next consider two copies of bosonic $q$-oscillators $\a_i,\ad_i,w_i=q^{N_i}$
which mutually commute. Then the Fock space is naturally spanned by vectors of
the form 
\begin{align}
|m,n\rangle = (\ad_1)^m (\ad_2)^n |0\rangle.
\end{align}
It is easy to see that under the identification
\begin{align}
&E = \ad_2 \a_1, && F = \ad_1 \a_2, && H = N_2-N_1,
\end{align}
the Fock space forms an infinite dimensional
$\mathcal{U}_q(\alg{sl}(2))$-representation. Moreover, the subspace
$\mathcal{F}_M = \mathrm{span}\{\,|m,M-m\rangle\;|\; m=0,\ldots,M\,\}$ is an
irreducible $\mathcal{U}_q(\alg{sl}(2))$-representation of dimension $M+1$. This
can be straightforwardly generalized to $\alg{sl}(n)$ and more generally, by
including fermionic oscillators, this space is extended to the representations
of $\alg{sl}(n|m)$ \cite{Chaichian}.


\paragraph{Representations of centrally extended $\mathcal{U}_q(\alg{su}(2|2))$.}

We will now construct the bound state representation for centrally extended
$\mathcal{U}_q(\alg{su}(2|2))$ in the $q$-oscillator language. We need to
consider two copies of $\alg{sl}(2)$, a bosonic and a fermionic one. Thus we
need four sets of $q$-oscillators $\a_i,\ad_i,w_i=q^{N_i}$, where the index
$i=1,2$ denotes bosonic oscillators and $i=3,4$ -- fermionic ones. Using these
we write
\begin{align}
&E_1 = \ad_2 \a_1, && F_1 = \ad_1 \a_2, && H_1 = N_2-N_1,\\
&E_2 = a~ \ad_4 \a_2 + b~ \ad_1 \a_3 && F_2 = c~ \ad_3 \a_1 + d~ \ad_2 \a_4 , &&
H_2 =-C +\frac{N_1+N_3-N_2-N_4}{2},\\
&E_3 = \ad_3 \a_4, && F_3 = \ad_4 \a_3, && H_3 = N_4-N_3,
\end{align}
where $C$ is central.
It is then straightforward to check that this set of generators forms a
representation of $\mathcal{U}_q(\alg{su}(2|2))$ on the Fock space when
restricting to the subspace of total particle number $M$ upon setting 
\begin{align}\label{eqn;CentralOsc}
  &ad = \frac{[C+{\textstyle \frac{M}{2}}]_q}{[M]_q}, && bc
=\frac{[C-{\textstyle \frac{M}{2}}]_{q}}{[M]_q}, && ab = \frac{\alg{P}}{[M]_q},
&& cd = \frac{\alg{K}}{[M]_q}.
\end{align}
In the above $\alg{K},\alg{P}$ correspond to the right hand side of the Serre
relations (\ref{eqn;quarticSerre}) following \cite{BK}. As a consequence, the
central charges satisfy the shortening condition
\begin{align}
[C]_q^2 -\alg{ PK} = \bigl[{\textstyle\frac{M}{2}}\bigr]_q^2.
\end{align}
Here the $q$-numbers are defined as
\begin{equation}
[k]_q = \frac{q^k-q^{-k}}{q-q^{-1}}. 
\end{equation}
This way of constructing representations of the centrally extended algebra
reminds us of the procedure used in, e.g.\ \cite{ALTlong}, where long
representations were be obtained by twisting $\alg{sl}(n|m)$ in a similar way.

In the $q\rightarrow 1$ limit the $q$-oscillators get reduced to regular
oscillators and their representations coincide with the superspace formalism
introduced in \cite{AFBound}. The identification is as follows
\begin{align}
&\a_{1,2} \leftrightarrow \frac{\partial}{\partial w_{1,2}}, && \ad_{1,2}
\leftrightarrow w_{1,2}, &&
\a_{3,4} \leftrightarrow \frac{\partial}{\partial \theta_{3,4}}, && \ad_{3,4}
\leftrightarrow \theta_{3,4}.
\end{align}
%


\paragraph{Parameterization and central elements.}

Introducing $V = q^C$ and $U$ as in \cite{BGM}, we rewrite
(\ref{eqn;CentralOsc}) as
\begin{align}
&ad = \frac{q^{\frac{M}{2}}V - q^{-\frac{M}{2}}V^{-1}}{q^M-q^{-M}}, && bc =
\frac{q^{-\frac{M}{2}}V - q^{\frac{M}{2}}V^{-1}}{q^M-q^{-M}},\el
&ab = \frac{g\alpha}{[M]_q}(1-U^2V^2), && cd =
\frac{g\alpha^{-1}}{[M]_q}(V^{-2}-U^{-2}).\label{param_2node}
\end{align}
which altogether leads to a constraint for $U$ and $V$,
\begin{equation}
 \frac{g^2}{[M]_q^2}(V^{-2}-U^{-2})(1-U^2V^2)=\frac{(V-q^M
V^{-1})(V-q^{-M}V^{-1})}{(q^M-q^{-M})^2} \;.
\end{equation}
This constraint agrees with the one in \cite{BGM} by identifying $q\rightarrow
q^M,\; g\rightarrow g/[M]_q$. The explicit parametrization of the labels
$a,b,c,d$ shall be given a bit further.


\subsection{Affine extension}

Next we want to consider the affine extension introduced in \cite{BGM}. Here we
will show that our representation allows an affine extension. Analogously to
\cite{BGM} we make the ansatz that the affine charges act as copies of
$E_2,F_2,H_2$. In other words, we set
\begin{align}
&E_4 = a_4~ \ad_4 \a_2 + b_4~ \ad_1 \a_3, 
  && F_4 = c_4~ \ad_3 \a_1 + d_4~ \ad_2 \a_4 , 
  && H_4 = -C_4 + \frac{N_1+N_3-N_2-N_4}{2}.
\end{align}
Checking all of the commutation relations is straightforward. Also, due to the
defining relations \eqref{param_2node}, the equivalent expressions for the
affine representation parameters are obtained
\begin{align}
&a_4 d_4 = \frac{q^{\frac{M}{2}}V_4 - q^{-\frac{M}{2}}V_4^{-1}}{q^M-q^{-M}}, 
  && b_4 c_4 = \frac{q^{-\frac{M}{2}}V_4 -
q^{\frac{M}{2}}V_4^{-1}}{q^M-q^{-M}},\el
&a_4 b_4 = \frac{g_4 \alpha_4}{[M]_q}(1-U_4^2 V_4^2), 
  && c_4 d_4 = \frac{g_4
\alpha_4^{-1}}{[M]_q}(V_4^{-2}-U_4^{-2}).\label{param_4node}
\end{align}
However the commutators between the generators $E_2$ and $E_4$ and also between
$F_2$ and $F_4$ induce relations between $a_2,\,a_4$, etc. These are found to be
\begin{align}
a_2 d_4 &= \frac{\tilde{g}\tilde{\alpha}^{-1}}{[M]_q} (q^{\frac{M}{2}}U_2
U_4^{-1} V_2 - q^{-\frac{M}{2}}V_4^{-1}),
  && b_2 c_4 = \frac{\tilde{g}\tilde{\alpha}^{-1}}{[M]_q} (q^{-\frac{M}{2}} U_2
U_4^{-1} V_2 - q^{\frac{M}{2}}V_4^{-1}),\el
c_2 b_4 &= \frac{\tilde{g}\tilde{\alpha}}{[M]_q} (q^{\frac{M}{2}}V_2^{-1} -
q^{-\frac{M}{2}}U_2^{-1} U_4 V_2),
  && d_2 a_4 = \frac{\tilde{g}\tilde{\alpha}}{[M]_q } (q^{-\frac{M}{2}}V_2^{-1}
- q^{\frac{M}{2}} U_2^{-1} U_4 V_2),\label{rel24}
\end{align}
and agree with \cite{BGM} upon sending $q\rightarrow
q^M,\,\tilde{g}\rightarrow\frac{\tilde{g}}{[M]_q}$, as in the non-affine case.
The tilded $\tilde{g},\; \tilde{\alpha}$ are not independent but constrained
parameters; thus there are 12 constraints for 12 parameters
$\{a_k,b_k,c_k,d_k,U_k,V_k\}$.


\paragraph{Hopf algebra and variables.}

The Hopf algebra structure is just as previously discussed in Section
\ref{sec;AffineAlgebra}. Here we will introduce Zhukowksy variables that will
parameterize the representation labels $\{a_k,b_k,c_k,d_k\}$ and central
elements $U_k,V_k$ for the bound-state representation. Following \cite{BGM} we
choose
\begin{equation}
g_2=g_4=g, \qquad \alpha_2=\alpha_4\,\tilde\alpha^{-2}=\alpha, \qquad \tilde
g^2=\frac{g^2}{1-g^2(q-q^{-1})^2}.\label{par24}
\end{equation}
Note that the powers of $q$ in the expressions above are $1$ and not $M$ because
$g^2(q-q^{-1})^2$ is invariant under the bound state map $(g,\,q) \mapsto
(g/[M]_q,\,q^M)$, thus these equations are identical to the ones for the
fundamental representation.

Also, there is a relation between the central elements of the algebra,
\begin{equation}
 U_4=\pm U_2^{-1}, \qquad V_4 = \pm V_2^{-1},
\end{equation}
that are called the two-parameter family of the representation \cite{BGM}. 
We shall be using the \textit{plus} relation in our calculations.

The mass-shell constraint (multiplet shortening condition) obtained from the
expressions (\ref{param_2node}) and (\ref{param_4node}) reads as
\begin{equation}
 (a_k d_k - q^M b_k c_k) (a_k d_k - q^{-M} b_k c_k) = 1,
\end{equation}
and holds independently for $k=2,4$. In terms of the conventional $x^\pm$
parametrization it becomes
\begin{equation}
 \frac{1}{q^{M}}\left(x^+ + \frac{1}{x^+}\right)-q^M\left(x^- +
\frac{1}{x^-}\right) = \left(q^M-\frac{1}{q^M}\right)
\left(\xi+\frac{1}{\xi}\right),\label{ms}
\end{equation}
where $\xi = -i \tilde g (q-q^{-1})$. One can further introduce a function
$\zeta(x)$
\begin{equation}
\zeta (x) =-\frac{x + 1/x + \xi + 1/\xi}{\xi-1/\xi},
\end{equation}
in terms of which (\ref{ms}) becomes $q^{-M}\zeta(x^{+}) = q^M\zeta(x^{-})$.
This parametrization leads to the following expressions of the labels
$a_k,b_k,c_k,d_k$ of a `canonical form':
\begin{align}
a_k &= \sqrt{\frac{g}{[M]_q}}\gamma_k,
  && b_k =
\sqrt{\frac{g}{[M]_q}}\frac{\alpha_k}{\gamma_k}\frac{x^-_k-x^+_k}{x^-_k},\el
c_k &= \sqrt{\frac{g}{[M]_q}}\frac{\gamma_k}{V_k\, \alpha_k}\frac{i\,
q^{\frac{M}{2}}\tilde{g}}{g (x^+_k + \xi)}, 
  && d_k = \sqrt{\frac{g}{[M]_q}} \frac{V_k\, \tilde{g}\, q^{\frac{M}{2}}}
{i\,g\,\gamma_k} \frac{x^+_k - x^-_k}{\xi x^+_k + 1},
\end{align}
where the central charges are
\begin{align}
&U^2_k = \frac{1}{q^M} \frac{x^+_k + \xi}{x^-_k + \xi} =  q^M
\frac{x^+_k}{x^-_k}\frac{\xi x^-_k + 1}{\xi x^+_k + 1}, 
  && V^2_k = \frac{1}{q^M} \frac{\xi x^+_k + 1}{\xi x^-_k + 1} = q^M
\frac{x^+_k}{x^-_k}\frac{x^-_k + \xi}{x^+_k + \xi},
\end{align}
and the relations between $x^{\pm}_2,\;\gamma_2$ and $x^{\pm}_4,\;\gamma_4$ are
constrained by (\ref{rel24}) to be
\begin{align}
\label{map24}
x_2^{\pm} &= x^{\pm}, 
 & x_4^{\pm} &= \frac{1}{x^{\pm}}, 
 & \gamma_2 &= \gamma, 
 & \gamma_4 &=\frac{i\tilde\alpha \gamma}{x^+}.
\end{align}
The relation between normalization coefficients $\alpha_2$ and $\alpha_4$ was
given in (\ref{par24}).
Finally, the convenient multiplicative evaluation parameter $z$
for the bound state representation is
\begin{equation}
z=q^{-M}\zeta(x^{+}) = q^M\,\zeta(x^{-}). 
\label{eval1}
\end{equation}
%


\subsection{Summary}

For the convenience of the reader we want to summarize all expressions that will
be used in the consequent calculations of the bound state S-matrix. We will
slightly change the notation for parameters related to the fermionic nodes. We
rename the representation parameters and the central elements of the algebra as
\begin{align}
 (a_2,b_2,c_2,d_2,U_2,V_2) \rightarrow (a,b,c,d,U,V), \el
 (a_4,b_4,c_4,d_4,U_4,V_4) \rightarrow (\tilde a,\tilde b,\tilde c,\tilde
d,\tU,\tV),
\end{align}
in order to reserve the subscript position for denoting the momentum
dependence, \textit{i.e.} $a_1 := a(p_1)$. 
in order to reserve the subscript position for discriminating states living in
different tensor spaces. We will also give some relations that we found to be very useful.


\paragraph{Explicit representation.} 
The bound state representation is defined as 
\begin{align}
\ket{m,n,k,l} = (\ad_3)^m(\ad_4)^n(\ad_1)^k(\ad_2)^l\,\ket{0}.
\end{align}
The total number of excitations is $k+l+m+n = M$. The triple corresponding to
the bosonic $\alg{sl}(2)$ is given by
\begin{align}
&H_1 |m,n,k,l\rangle = (l-k) |m,n,k,l\rangle, && \el
&E_1 |m,n,k,l\rangle = [k]_q\, |m,n,k-1,l+1\rangle, && F_1 |m,n,k,l\rangle=
[l]_q\, |m,n,k+1,l-1\rangle.
\end{align}
The fermionic part is
\begin{align}
&H_3 |m,n,k,l\rangle = (n-m) |m,n,k,l\rangle,  \el
&E_3 |m,n,k,l\rangle = |m+1,n-1,k,l\rangle, && F_3 |m,n,k,l\rangle =
|m-1,n+1,k,l\rangle.
\end{align}
The action of the supercharges is given by
\begin{align}
H_2 |m,n,k,l\rangle =&~ -\left\{C-\frac{k-l+m-n}{2}\right\} |m,n,k,l\rangle, 
\el
E_2 |m,n,k,l\rangle =&~ a~(-1)^{m}[l]_q\, |m,n+1,k,l-1\rangle +  b~
|m-1,n,k+1,l\rangle, \el
F_2 |m,n,k,l\rangle =&~ c~[k]_q\, |m+1,n,k-1,l\rangle + d~(-1)^{m}\,
|m,n-1,k,l+1\rangle.
\end{align}
The parameters $a,b,c,d$ are related to the central charges via
(\ref{eqn;CentralOsc}). The affine charges are defined exactly in the same way,
\begin{align}
H_4 |m,n,k,l\rangle =&~ -\left\{\widetilde{C}-\frac{k-l+m-n}{2}\right\}
|m,n,k,l\rangle,  \el
E_4 |m,n,k,l\rangle =&~ \tilde{a}~(-1)^{m}[l]_q |m,n+1,k,l-1\rangle + 
\tilde{b}~ |m-1,n,k+1,l\rangle, \el
F_4 |m,n,k,l\rangle =&~ \tilde{c}~[k]_q\, |m+1,n,k-1,l\rangle +
\tilde{d}~(-1)^{m}\, |m,n-1,k,l+1\rangle.
\end{align}
The representation labels $a,b,c,d$ are given by
\begin{align}
a &= \sqrt{\frac{g}{[M]_q}}\gamma,   && 
b  = \sqrt{\frac{g}{[M]_q}}\frac{\alpha}{\gamma}\frac{x^--x^+}{x^-}, \el
c &=
\sqrt{\frac{g}{[M]_q}}\frac{\gamma}{\alpha\,V}\frac{i\,q^{\frac{M}{2}}\tilde{g}}
{g (x^+ + \xi)}, && 
d  = \sqrt{\frac{g}{[M]_q}} \frac{\tilde{g}\,
q^{\frac{M}{2}}V}{i\,g\,\gamma}\frac{x^+ - x^-}{\xi x^+ + 1}, \label{abcd}
\end{align}
and the affine parameters $\tilde a, \tilde b, \tilde c, \tilde d$ are acquired
by replacing $V\to\tV =  V^{-1}$, $\gamma\to\frac{i\tilde{\alpha}\gamma}{x^+}$,
$\alpha\to\,\alpha\,\tilde\alpha^2$ and $x^{\pm}\to\frac{1}{x^{\pm}}$; the
corresponding central elements are given by $V = q^C$, $\tV =
q^{\widetilde{C}}$.\\


\paragraph{Useful relations.}
The evaluation parameter $z$ may be expressed explicitly in terms of $x^{\pm}$
parametrization as
\begin{align}\label{zrel12}
z\,(q-q^{-1})(\xi-\xi^{-1}) = -\frac{1}{[M]_q} \Bigl(x^+ - x^-
+\frac{1}{x^+}-\frac{1}{x^-}\Bigr). 
\end{align} 
Then using the identity
\begin{align}
\xi-\xi^{-1} = \frac{\tilde g}{i(q-q^{-1})g^2},
\end{align}
one can further show that it is related to the representation labels
(\ref{abcd}) and their affine partners in a very nice way,
\begin{align}
z=\frac{g}{\tl g\,\alpha\,\tl\alpha}(a\tl b-b\tl a), \qquad
  \frac{1}{z}=\frac{g\,\alpha\,\tl\alpha}{\tl g}(c\tl d-d\tl c),
\end{align}
while the consistency conditions (\ref{rel24}) give
\begin{align}\label{zrel3}
z=\frac{1-U^2V^2}{V^2-U^2} = \frac{1-\tU^2\tV^2}{\tV^2-\tU^2}\,.
\end{align}
%


\paragraph{Rational limit.}

The rational limit is usually obtained by substituting $q=1+h$ and then finding
the $h\to0$ limit. Thus by defining the evaluation parameter \eqref{eval1} as
$z=q^{-2u}$ we can expand it in series of $h$ as \cite{BGM}
\begin{align}\label{zh}
z=1-2hu+\mathcal{O}(h^2), \qquad \text{where}\qquad
u=\frac{ig}{2}(x^++x^-)(1+1/x^+x^-).
\end{align}
It is noted that the $x^\pm$ parameters in \eqref{zh} satisfies the leading
order of the following relation 
which is stemming from the mass-shell constraint \eqref{ms} in the $h\to0$
limit, 
\begin{align}
x^{+}+\frac{1}{x^{+}}-x^--\frac{1}{x^{-}} = \frac{i M}{g}+2hM
u+\mathcal{O}(h^2).
\end{align}
In fact, this is consistent with the rational constraint for $x^\pm$ parameters
\cite{ALT}. 
Finally, it would be important to see how the representation parameters reduce
in the rational limit. 
The representation labels \eqref{abcd} in the $q\to1$ limit reduce to the usual
(undeformed) labels $(a,b,c,d)$ of \cite{ALT}. 
On the other hand, the affine parameters are related to the non-affine ones
$(\tl a,\tl b,\tl c,\tl d)$ through \cite{BGM} 
\begin{align}\label{t4t2}
M\widetilde{T}= \begin{pmatrix}z^{-1}&0\\0&1\end{pmatrix} T
\begin{pmatrix}w^{-1}&0\\0&wz\end{pmatrix}
\quad\text{with}\quad 
M=\begin{pmatrix}0&\alpha\tilde\alpha
\\-\alpha^{-1}\tilde\alpha^{-1}&0\end{pmatrix}, \quad 
T=\begin{pmatrix}a&-b\\-c&d\end{pmatrix}
\end{align}
where $z$ is the evaluation parameter given in \eqref{eval1}, \eqref{zrel3} and
$w$ is defined by 
\begin{align}
w=\frac{\tilde g V}{g q^{1/2}}\frac{qU^2-1}{V^2U^2-1}=\frac{g q^{1/2}}{\tilde g
V}\frac{U^2-V^2}{U^2-q}.  
\end{align}
Since the central elements specialize to $(U,V) \to
(\sqrt{\frac{x^{+}}{x^{-}}},1)$ in the limit $q\to1$, 
it is easy to see that the matrix relation \eqref{t4t2} reduces the following
simple form, 
\begin{align}
M\widetilde{T}=T.  
\end{align}
%


\section{The S-matrix}\label{sec:Smat}

We shall consider the bound state S-matrix which is an intertwining matrix of
the tensor space furnished by the vectors
\begin{align}\label{bstates}
|m_1,n_1,k_1,l_1\rangle\otimes|m_2,n_2,k_2,l_2\rangle.
\end{align}
Here $0 \leq m_1, n_1, m_2, n_2 \leq 1$ and $k_1, l_1, k_2, l_2 \geq 0$ denote
the numbers of fermionic and bosonic excitations respectively with the bound
state number $M_i$ being the total number of excitations $M_i = m_i + n_i + k_i
+ l_i$. Thus the S-matrix is the automorphism of the quantum deformed tensor
space and is required to be invariant under the coproducts of the affine algebra
$\afQ$,
\begin{align}\label{Sinv}
\S\,\Delta(J) = \Delta^{op}(J)\, \S, \qquad\text{for any}\qquad J\in\afQ.
\end{align}
We normalize the S-matrix in such a way that the state
$|0,0,0,M_1\rangle\otimes|0,0,0,M_2\rangle$ is invariant under the scattering.
Therefore we will denote the state
\begin{equation}\label{vacuum}
 |0\rangle = |0,0,0,M_1\rangle\otimes|0,0,0,M_2\rangle,
\end{equation}
as the vacuum state. 

The invariance under bosonic symmetries $\DH_1$ and $\DH_3$ requires the total
number of fermions and the total number of fermions of one type\footnote{Note
that a bosonic excitation may be interpreted as a combined excitation of two
fermions of different type.}
\begin{align}
N_f &= m_1 + m_2 + n_1 + n_2 + 2 l_1 + 2 l_2,\el
N_{f_3} &= m_1 + m_2 + l_1 + l_2.\label{Nf}
\end{align}
to be conserved. This conservation divides the space \eqref{bstates} into five
types of invariant subspaces of the S-matrix:
\begin{description}
\item[I$\quad$]  $|0,1,k_1,l_1\rangle\otimes|0,1,k_2,l_2\rangle$,
\item[Ib$\;\;$] $|1,0,k_1,l_1\rangle\otimes|1,0,k_2,l_2\rangle$,
\item[II$\;\;$]
$\{|0,0,k_1,l_1\rangle\otimes|0,1,k_2,l_2\rangle,|1,1,k_1,l_1\rangle\otimes|0,1,
k_2,l_2\rangle,\\
\quad|0,1,k_1,l_1\rangle\otimes|0,0,k_2,l_2\rangle,|0,1,k_1,l_1\rangle\otimes|1,
1,k_2,l_2\rangle\}$,
\item[IIb]
$\{|0,0,k_1,l_1\rangle\otimes|1,0,k_2,l_2\rangle,|1,1,k_1,l_1\rangle\otimes|1,0,
k_2,l_2\rangle,\\
\quad|1,0,k_1,l_1\rangle\otimes|0,0,k_2,l_2\rangle,|1,0,k_1,l_1\rangle\otimes|1,
1,k_2,l_2\rangle\}$,
\item[III]
$\{|0,0,k_1,l_1\rangle\otimes|0,0,k_2,l_2\rangle,|0,0,k_1,l_1\rangle\otimes|1,1,
k_2,l_2\rangle,|1,1,k_1,l_1\rangle\otimes|0,0,k_2,l_2\rangle,\\
\quad|1,1,k_1,l_1\rangle\otimes|1,1,k_2,l_2\rangle,|0,1,k_1,l_1\rangle\otimes|1,
0,k_2,l_2\rangle,|1,0,k_1,l_1\rangle\otimes|0,1,k_2,l_2\rangle\}$.
\end{description}
Subspaces I, Ib and II, IIb are isomorphic, hence we need to find the S-matrix
for one of the isomorphic subspaces only. 
In the following we will consider the scattering in the subspaces I, II and III
only. 

\begin{figure}\centering
\epsfig{file=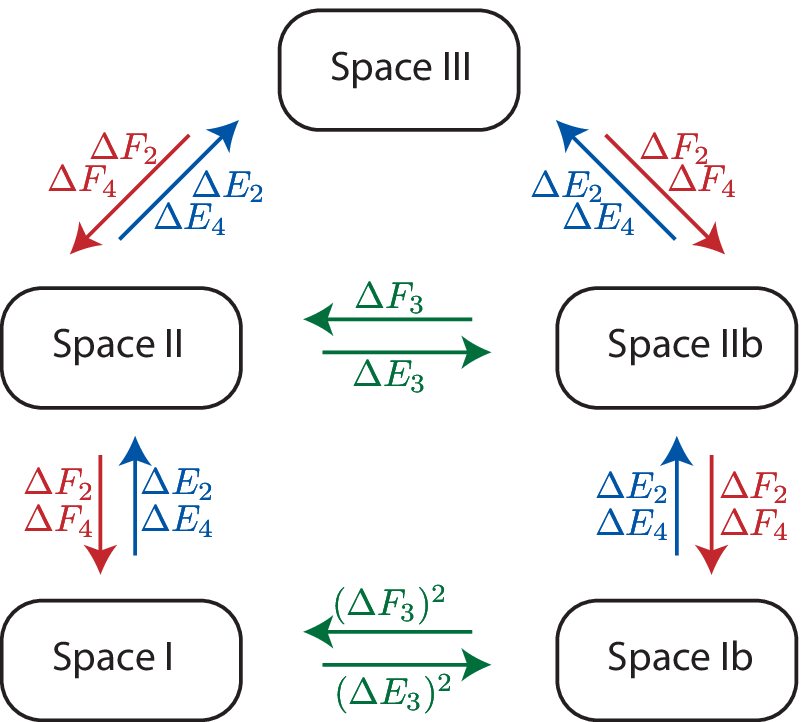,width=7cm}
\caption{The invariant subspaces of the S-matrix and the algebraic relations
between them.}
\label{fig:Subspaces}
\end{figure}

The invariant subspaces differ by the numbers $N_{f,f_3}$. By considering the
action of the algebra charges it is easy to see that the different subspaces are
related to each other in the way shown in figure \ref{fig:Subspaces}.

Finally we want to give a remark on our choice of the basis. The $q$-oscillator
basis we are considering is orthogonal, but not orthonormal,
\begin{align}
\langle m',n',k',l'|m,n,k,l\rangle = \frac{1}{[k]!\,[l]!}
\,\delta_{m,m'}\delta_{n,n'}\delta_{k,k'}\delta_{l,l'},
\end{align}
where $[n]!=[n]_q[n-1]_q\cdots[1]_q$ is the quantum factorial. 
We shall choose the normalization for the \textit{bra} vectors to be
\begin{align}
\langle m,n,k,l| := \frac{1}{[k]!\,[l]!}\,|m,n,k,l\rangle^\dagger.
\end{align}
which helps us to normalize the scalar product to unity and avoid the appearance
of unpleasant numerical factors of the form $\bigl([k]![l]!\bigr)^{-1/2}$ in the
derivations. The price we have to pay for this choice of the basis is
that, for real $q$, the inverse of the S-matrix is related to the
Hermitian conjugate only up to a basis transformation. For complex $q$ this
property is not valid even for the fundamental representation \cite{BK}. 

\noindent For further convenience we introduce these shorthands 
\begin{align}
M &= M_1+M_2, \qquad \delta M = M_1-M_2,\qquad K=k_1+k_2, \qquad\delta
K=k_1-k_2, \el
\bar{k}_i &= M_i-k_i-1, \quad\! \delta k_i = \bar{k_i}-k_i=M_i-2k_i-1, \quad
z_{12}=z_1/z_2, \quad \delta u=u_1-u_2. 
\end{align}
%


\subsection{Scattering in subspace I}\label{sI}

The conserved fermionic numbers \eqref{Nf} for the subspace I are $N_f = 2K+2$
and $N_{f_3} = K + 2$. Thus for the fixed $K$ ($0 \leq K \leq M_1 + M_2 - 2$)
the dimension of the space is $K+1$ and the states in this space are defined as
\begin{align}
\stateA{k_1,k_2} &= |0,1,k_1,M_1-k_1-1\rangle\otimes|0,1,k_2,M_2-k_2-1\rangle. 
\end{align}
We start by considering the highest weight state (the state with $k_1=k_2=0$).
The invariance under $\DH_1$ and $\DH_3$ requires it to be an eigenstate of the
S-matrix,
\begin{align}\label{SD}
\S\,|0,0\rangle^{\rm I} = \mathscr{D}\,|0,0\rangle^{\rm I}.
\end{align}
Let us compute $\mathscr{D}$. First, we construct the highest weight state by
acting with the combination $\DE_2\DE_4$ on the vacuum state \eqref{vacuum} (we
use the notation $a_i\equiv a(p_i)$ etc.)
\begin{align}
&\DE_2 \DE_4 \, |0\rangle
=q^{\frac{M_1}{2}}[M_1]_q [M_2]_q \,(a_1\tilde{a}_2\, \tU_1\tV_1 -
a_2\tilde{a}_1\,U_1 V_1)\,|0,0\rangle^{\rm I}.
\end{align}
This construction let us to rewrite (\ref{SD}) as 
\begin{align}
\S\,|0,0\rangle^{\rm I} &= \frac{\S\, \DE_2 \DE_4 }{q^{\frac{M_1}{2}}[M_1]_q
[M_2]_q \,(a_1\tilde{a}_2\, \tU_1\tV_1 - a_2\tilde{a}_1\,V_1 U_1)}\,|0
\rangle\el
  &= \frac{\DOE_2 \DOE_4 \,\S}{q^{\frac{M_1}{2}}[M_1]_q [M_2]_q
\,(a_1\tilde{a}_2\, \tU_1\tV_1 - a_2\tilde{a}_1\,V_1 U_1)}\,|0\rangle \el
  &= -q^{\frac{M_2-M_1}{2}}\frac{a_2\tilde{a}_1\, \tU_2\tV_2 -
a_1\tilde{a}_2\,V_2 U_2}{a_1\tilde{a}_2\, \tU_1\tV_1 - a_2\tilde{a}_1\,V_1
U_1}\,|0,0\rangle^{\rm I}, \label{findD}
\end{align}
where we have used the invariance condition \eqref{Sinv} when going from the
first to the second line. Comparing \eqref{findD} with \eqref{SD} we find
$\mathscr{D}$ to be
\begin{align}\label{eqn;D}
\mathscr{D} 
= -q^{\frac{M_2-M_1}{2}}\frac{a_2\tilde{a}_1\, \tU_2\tV_2 - a_1\tilde{a}_2\,V_2
U_2}{a_1\tilde{a}_2\, \tU_1\tV_1 - a_2\tilde{a}_1\,V_1 U_1}
= q^{-\delta M/2}\frac{U_2V_2}{U_1V_1}\frac{\xp - \xmm}{\xm - \xpp}. 
\end{align}
In the $q\to1$ limit this is the inverse of the result found in \cite{ALT} due
to the interchange of $\Delta$ and $\Delta^{op}$ with respect to the ones in
\cite{ALT}.\\


Next we define the action of the S-matrix on the subspace I to be
\begin{align}
\S \,\stateA{k_1,k_2} &= \sum^{K}_{n=0} \mathscr{X}^{k_1,k_2}_{n}
\stateA{n,K-n}.
\end{align}
The strategy for finding coefficients $\mathscr{X}^{k_1,k_2}_{n}$ will be based
on building the generic state $\stateA{k_1,k_2}$ by starting from the highest
weight state $\stateA{0,0}$. This allows us to relate
$\mathscr{X}^{k_1,k_2}_{n}$ with any $k_1$, $k_2$ and $n$ to the already known
coefficient $\mathscr{D}$. Thus we need to construct $k_1$- and $k_2$-raising
operators. We start from inspecting the action of the coproduct of the bosonic
charge $F_1$ giving
\begin{align}
&\DF_1 \stateA{k_1,k_2}  = [\bar{k}_1]_q\, q^{\delta k_2}\stateA{k_1+1,k_2}  +
[\bar{k}_2]_q\,\stateA{k_1,k_2+1},
\end{align}
and
\begin{align}
&\DOF_1 \stateA{k_1,k_2}  = [\bar{k}_1]_q\,\stateA{k_1+1,k_2} + [\bar{k}_2]_q\,
q^{\delta k_1}\stateA{k_1,k_2+1}.
\end{align}
These coproducts do not have the desired properties we want, but are very close.
However, with the help of $E_2$, $E_3$ and $E_4$ we can construct a new charge
with a similar action, 
\begin{align}
&\hat{F}_1 = \frac{g}{\tilde g\,\alpha\,\tilde{\alpha}} 
\,\{E_2,[E_4,E_3]\}.\label{newF1}
\end{align}
We call this new charge `the affine partner' of the raising charge $F_1$. The
action of $\hat{F}_1 $ on the state of the form $|0,1,k,l\rangle$ is
\begin{align}
\hat{F}_1 |0,1,k,l\rangle &= z\,[l]_q\, |0,1,k+1,l-1\rangle,
\end{align}
where we have used (\ref{zrel12}) implicitly\footnote{For the consistency of the
algebra we also give a definition of the `affine lowering charge' $\hat{E}_1$:
\begin{align}
\hat{E}_1 = \frac{g\,\alpha\,\tl\alpha}{\tilde g} 
\,\{F_2,[F_4,F_3]\},\qquad \hat{E}_1 |0,1,k,l\rangle =
\frac{[k]_q}{z}\,|0,1,k-1,l+1\rangle.
\end{align}
}. 
Then it is straightforward to see that the new affine raising charge acts on
generic states in subspace I as 
\begin{align}
\DHF_1\,\stateA{k_1,k_2} &= z_1\,[\bar{k}_1]_q\, \stateA{k_1+1,k_2} +
z_2\,q^{\delta k_1}\,[\bar{k}_2]_q\,\stateA{k_1,k_2+1}.
\end{align}
And the action of $\Delta^{\!op}\hat{F}_1$ is
\begin{align}
\Delta^{\!op}\hat{F}_1\,\stateA{k_1,k_2} &= z_1\,q^{\delta
k_2}\,[\bar{k}_1]_q\,\stateA{k_1+1,k_2} + z_2\,[\bar{k}_2]_q\,
\stateA{k_1,k_2+1}.
\end{align}
By combining $\DHF_1$ with $\DF_1$ we obtain composite operators having the
action of the desired form -- raising $k_1$ and $k_2$ separately:
\begin{align}
\stateA{k_1+1,k_2} &= \frac{1}{[\bar{k}_1]_q}\frac{\DHF_1 - z_2\,q^{\delta k_1}
\DF_1}
  {z_1- z_2\, q^{\delta k_1+\delta k_2}} \,\stateA{k_1,k_2},\\
\stateA{k_1,k_2+1} &= \frac{1}{[\bar{k}_2]_q} \frac{z_1\,\DF_1 - q^{\delta
k_2}\DHF_1}
  {z_1 - z_2\, q^{\delta k_1+\delta k_2}} \,\stateA{k_1,k_2}.
\end{align} 
Then by induction we find that the generic state $\stateA{k_1,k_2} $ may be
constructed as
\begin{align}
\stateA{k_1,k_2}  = \frac{\prod_{j_2=0}^{k_2-1}(z_1\, \DF_1 -q^{\delta j_2}
\DHF_1
)\prod_{i_1=0}^{k_1-1}(\DHF_1 - z_2\,q^{\delta i_1} \DF_1)}
{\prod_{i=1}^{k_1}[M_1-i]_q\prod_{j=1}^{k_2}[M_2-j]_q\prod_{j=1}^{k_1+k_2} (z_1
- z_2 q^{M-2j})}\,|0,0\rangle^{\rm I}.
\end{align}
Finding $\mathscr{X}^{k_1,k_2}_{n}$ is then straightforward. We only need to act
with the S-matrix on the expression above and sandwich with a {\it bra}-vector
as
\begin{align}
\mathscr{X}^{k_1,k_2}_{n} = \,\costateA{n,K-n}|~\S~\stateA{k_1,k_2}.
\end{align}
Performing similar steps as we did in (\ref{findD}) and employing the relations
\begin{align}
&(\Delta^{\!op}\hat{F}_1 - z_2\,q^{\delta k_1} \DOF_1)\, \stateA{n_1,n_2}  \el
& \qquad =[\bar{n}_2]_q\, z_2\, (1-q^{\delta k_1+\delta n_1}) \,
\stateA{n_1,n_2+1} + 
  [\bar{n}_1]_q\, (z_1\, q^{\delta n_2 }- z_2\, q^{\delta k_1}) \,
\stateA{n_1+1,n_2},\label{op1}\\
& (z_1\, \DOF_1 -q^{\delta k_2} \Delta^{\!op}\hat{F}_1)\, \stateA{n_1,n_2} \el 
& \qquad = [\bar{n}_1]_q\,z_1\,(1- q^{\delta n_2+\delta k_2})\,
\stateA{n_1+1,n_2} + 
  [\bar{n}_2]_q\,(z_1 q^{\delta n_1}- z_2 q^{\delta k_2})
\stateA{n_1,n_2+1},\label{op2}
\end{align}
we find the coefficients of the S-matrix in the subspace I to be
\begin{align}
\mathscr{X}^{k_1,k_2}_{n} &= \mathscr{D}\,
\frac{\prod_{i=1}^{n}[M_1-i]_q\prod_{j=1}^{K-n}[M_2-j]_q}{\prod_{i=1}^{k_1}[
M_1-i]_q\prod_{j=1}^{k_2}[M_2-j]_q}
\frac{1}{\prod_{l=1}^{K} (z_{12} - q^{M-2l})}\el
&\quad\times\sum_{m=0}^{k_1}\left(z_{12}^{n-m}q^{k_2 (n-m)-k_1 m-k_2^2}
{\scriptscriptstyle \left[\begin{array}{c} k_1 \\
m\end{array}\right]_{q}\left[\begin{array}{c} k_2 \\ n-m\end{array}\right]_{q}
}\right.\el
&\qquad\qquad\quad \times\prod_{p=0}^{m-1} (z_{12}\, q^{M_2+2p} -
q^{M_1})\prod_{p=1+m}^{k_1} (1-q^{2(M_1-p)})\el
&\qquad\qquad\quad \times \left. \prod_{p=1}^{n-m}(1-q^{2(M_2-K+n-p)})
\prod_{p=-m}^{k_2-n-1}(z_{12}\, q^{M_1+2p} - q^{M_2})\right),
\label{coefX}\end{align}
where $z_{12} = \frac{z_1}{z_2}$ and the $q$-binomials are defined as
\begin{align}
{\scriptscriptstyle \left[\begin{array}{c} a \\ b\end{array}\right]_q}\equiv
\frac{[a]_q!}{[b]_q![a-b]_q!}.
\end{align}
Apart from the prefactor $\mathscr{D}$, this expression only depends on the
quotient $z_{12}$ and on simple $q$-factors. The expression above has exactly
the form that one would expect to obtain by an educated guess relying on the one
given in \cite{ALT}.


\paragraph{Quantum $6j$-Symbol.}

The coefficients $\mathscr{X}^{k_1,k_2}_n$ of the bound state S-matrix may be
regarded as the coefficients which arise in the fusion rule of the irreducible
representations of $\mathcal{U}_q(su(2))$, thus it is expected that the
expression \eqref{coefX} is related to the quantum $6j$-symbol, which is the
$q$-deformation of $6j$-symbol and was first introduced in \cite{KR}. 

In order to see the relation with the quantum $6j$-symbol, we first rewrite
\eqref{coefX} in terms of quantum factorials.  
This can be done by introducing the notation $z_{12}=q^{-2\delta u}$ and using
the following identity several times,
\begin{align}
\frac{q^A-q^B}{q-q^{-1}}=q^{\frac{A+B}{2}}\left[\frac{A-B}{2}\right]_q.
\end{align} 
Secondary, we shift the index of summation $m$ to $M_1-2-m$. After some
computation, we obtain the following form,
\begin{align}
\mathscr{X}^{k_1,k_2}_{n} 
&=\mathscr{D}\, q^{(k_1-n)(k_2-n+\delta u+\frac{\delta M}{2})}
\frac{[M_2-k_2-1]!}{[M_1-n-1]!}
\frac{[\delta u+\frac{M}{2}-1-K]!}{[\delta u+\frac{M}{2}-1]!}
\el
&\times 
[k_1]![k_2]![\delta u+\tfrac{\delta M}{2}]! [\delta u-\tfrac{\delta
M}{2}-k_2+n+1]!
\el
&\times
\sum_{m\geq 0}[m+1]!\left(
[m-M_1+2+k_1]!\,
[m-M_1+2+n]!\,
[k_2-n+M_1-2-m]!
\right.\el
&\quad \times \left.
[m+\delta u-\tfrac{M}{2}+2]!\,
[\delta u+\tfrac{M}{2}-1-m]!\,
[M_1-2-m]!\,
[M-K-3-m]!\right)^{-1}.
\end{align}   
where the summation index $m$ runs over the non-negative integers such that all
arguments of the quantum factorials, which do not include $\delta u$, are
non-negative.
Finally, replacing the six variables $(M_1,M_2,k_1,k_2,n,\delta u)$ by the
appropriate combinations of $(j_1,j_2,j_3,j_4,j_5, j_6)$ as (see also \cite{ALT}),  
\begin{align}
j_1&= \tfrac{1}{2} (K - n + \tfrac{\delta M}{2} + \delta u), & 
j_4&= \tfrac{1}{2} (\tfrac{\delta M}{2} - 1 + k_2 - \delta u),\el  
j_2&= \tfrac{1}{2} (\tfrac{M}{2}-2-k_2 - \delta u), & 
j_5&= \tfrac{1}{2} (\tfrac{M}{2} - 1 - K + n + \delta u),\el  
j_3&= \tfrac{1}{2} (M_1 - 2 - k_1 - n), &  
j_6&= \tfrac{1}{2} (M_2 - 1),
\end{align} 
we have found that the expression \eqref{coefX} obtains a quite elegant form
\begin{align}
\label{X6j}
\mathscr{X}^{k_1,k_2}_{n} 
&= \mathscr{D}\, (-1)^{j_1-j_3-j_4+2j_5+j_6}q^{(j_1 - j_2 + j_3)(j_1 + j_2 - j_4
- j_5)}
\frac{[j_1 + j_2 - j_3]!}{[1 + j_1 + j_2 + j_3]!}\,
\frac{[j_1 + j_5 - j_6]!}{[j_1 + j_5 + j_6]!}
\el
&\times
[j_3 - j_4 + j_5 ]!\, [j_3 + j_4 - j_5]!\, [j_2 - j_4 + j_6]!\, [-j_2 + j_4 +
j_6]! \;
{\scriptscriptstyle \left|\begin{array}{ccc} j_1&j_2&j_3 \\
j_4&j_5&j_6\end{array}\right| },
\end{align}   
where we have defined the rescaled quantum $6j$-symbol by  
\begin{align}
{\scriptscriptstyle \left|\begin{array}{ccc} j_1&j_2&j_3 \\
j_4&j_5&j_6\end{array}\right| }
&=\sum_{m\geq 0}(-1)^m[m+1]! 
\bigl([j_{1245}- m]!\,[j_{1346}- m]!\,[j_{2356}-m]!
\el
&\qquad\quad\times
[m-j_{123}]!\,[m-j_{345}]!\,[m-j_{246}]!\,[m-j_{156}]!
\bigr)^{-1}.
\end{align}
Here we have used the bookkeeping notations $j_{abc}=j_a+j_b+j_c$ and
$j_{abcd}=j_a+j_b+j_c+j_d$. The above expression is related with the quantum
$6j$-symbol introduced in \cite{KR} as
\begin{align}
{\scriptscriptstyle \left\{\begin{array}{ccc} j_1&j_2&j_3 \\
j_4&j_5&j_6\end{array}\right\} }
&=\sqrt{2j_3-1}\sqrt{2j_6-1}\, (-1)^{-j_1-j_2+2j_3+j_4+j_5}
\el
&\quad\times
\mathrm{\Delta}(j_1,j_2,j_3)\mathrm{\Delta}(j_1,j_5,j_6)\mathrm{\Delta}(j_2,j_4,
j_6)\mathrm{\Delta}(j_3,j_4,j_5)
{\scriptscriptstyle \left|\begin{array}{ccc} j_1&j_2&j_3 \\
j_4&j_5&j_6\end{array}\right| },
\end{align}
where the triangle coefficient $\mathrm{\Delta}(a,b,c)$ is defined to be 
\begin{align}
\mathrm{\Delta}(a,b,c)=\left(\frac{[a+b-c]!\,[b+c-a]!\,[c+a-b]!}{[1+a+b+c]!}
\right)^{1/2}.
\end{align}
%


\paragraph{Rational Limit.}

In order to find the rational limit of the matrix $\mathscr{X}$ \eqref{coefX} we
first use the expansion \eqref{zh} for the spectral parameter $z$. This leads to
\begin{align}
\mathscr{X}^{k_1,k_2}_{n} 
&= \mathscr{D}\,
\frac{\prod_{i=1}^{n}[M_1-i]_q\prod_{j=1}^{K-n}[M_2-j]_q}{\prod_{i=1}^{k_1}[
M_1-i]_q\prod_{j=1}^{k_2}[M_2-j]_q}
\frac{1}{\prod_{l=1}^{K} (z_{12}^{1/2}[\delta u]_q +
q^{M/2-l}[\frac{M}{2}-l]_q)}\el
&\quad\times\sum_{m=0}^{k_1}\left(z_{12}^{n-m}q^{k_2 (n-m)-k_1 m-k_2^2}
{\scriptscriptstyle \left[\begin{array}{c} k_1 \\
m\end{array}\right]_{q}\left[\begin{array}{c} k_2 \\ n-m\end{array}\right]_{q}
}\right.\el
&\qquad\qquad\quad \times
\prod_{p=0}^{m-1} \left(z_{12}^{1/2}\, q^{M_2/2+p}\left[\delta u
-\frac{M_2}{2}-p\right]_q + q^{M_1/2}\left[\frac{M_1}{2}\right]_q\right)\el
&\qquad\qquad\quad \times
\prod_{p=-m}^{k_2-n-1}\left(z_{12}^{1/2}\,q^{M_1/2+p}\left[\delta
u-\frac{M_1}{2}-p\right]_q 
+ q^{M_2/2}\left[\frac{M_2}{2}\right]_q\right)
\el
&\qquad\qquad\quad \times \left. 
\prod_{p=1+m}^{k_1} q^{M_1-p}[M_1-p]_q 
\prod_{p=1}^{n-m}q^{M_2-K+n-p}[M_2-K+n-p]_q 
\right),\label{Xrat1}
\end{align} 
where $\delta u=u_1-u_2$. 
Now we are ready to find $q\to1$ limit. The $q$-numbers $[x]_q$ coalesce to $x$,
thus \eqref{Xrat1} becomes
\begin{align}
\mathscr{X}^{k_1,k_2}_{n} 
&= \mathscr{D}\,
\frac{\prod_{i=1}^{n}(M_1-i)\prod_{j=1}^{K-n}(M_2-j)}{\prod_{i=1}^{k_1}
(M_1-i)\prod_{j=1}^{k_2}(M_2-j)}
\frac{1}{\prod_{l=1}^{K} (\delta u+ \frac{M}{2}-l)}\el
&\quad\times\sum_{m=0}^{k_1}\left(
{\scriptscriptstyle \left(\begin{array}{c} k_1 \\ m\end{array}\right)
\left(\begin{array}{c} k_2 \\ n-m\end{array}\right)
}\right.
\prod_{p=0}^{m-1} \left(\delta u +\frac{\dM}{2}-p\right)
\prod_{p=-m}^{k_2-n-1} \left(\delta u -\frac{\dM}{2}-p\right)
\el
&\qquad\qquad\quad \times \left. 
\prod_{p=1+m}^{k_1} (M_1-p)
\prod_{p=1}^{n-m}(M_2-K+n-p) 
\right).
\end{align}
This result coincides exactly with the expression obtained in \cite{ALT}%
\footnote{The normalization of the evaluation parameter is slightly different in
here, $u_{\text{here}}=-2u$\tiny{\cite{ALT}}.} 


\paragraph{Classical Limit.}

It is also important to find the classical limit $g\to\infty$ of \eqref{coefX}. 
This limit corresponds to the case `T(h)' in the analysis of the classical
algebra \cite{Bcl}, 
where the deformation parameter $q$ is expanded as 
\begin{align}
q=1+\frac{h}{2g}+\mathcal{O}(g^{-2}),
\end{align}
and the $x^\pm$ parameters become 
\begin{align}
x^\pm=x\left[1\pm\frac{hM}{2g}\frac{(x+\tilde h)(1+1/x\tilde
h)}{x-x^{-1}}+\mathcal{O}(g^{-2})\right],
\quad\text{where}\quad 
\tilde h=-\frac{ih}{\sqrt{1-h^2}}\;.
\end{align}
The above expressions are compatible with the constraint \eqref{ms} up to a
given order.  
Since $\xi\to \tilde h$ and $x^\pm\to x$ in the classical limit, 
it is easy to see that the evaluation parameter $z$ reduces to\footnote{%
The classical evaluation parameter given in \cite{Bcl} is related with ours
as $z^{cl}_{\text{[34]}}=(z^{cl}_{\text{here}})^{-1}$ 
and the classical parameter is $x_{\text{[34]}}=-ih{\tilde
h}^{-1}(x_{\text{here}}+\tilde h)$.} 
\begin{align}\label{clev}
z=-\frac{(x+\tilde h)(1+1/x\tilde h)}{\tilde h-{\tilde
h}^{-1}}=-\frac{C+D}{C-D}\,,
\end{align}
where elements $C$ and $D$ are the classical limits of $U=q^D$ and $V=q^C$
respectively, 
and are given by 
\begin{align}
D=\tfrac{1}{2}(z+1)\,\tl q, \qquad C=\tfrac{1}{2}(z-1)\,\tl q,
\qquad\text{where}\qquad
\tilde q=-M\frac{{\tilde h}-{\tilde h}^{-1}}{x-x^{-1}}.
\end{align}
With these preliminaries, we find the classical limit of \eqref{coefX} to be 
\begin{align}
\mathscr{X}^{k_1,k_2}_{n}
&\sim (1+\mathscr{D}_{cl})\, 
\frac{\prod_{i=1}^{n}(M_1-i)\prod_{j=1}^{K-n}(M_2-j)}{\prod_{i=1}^{k_1}
(M_1-i)\prod_{j=1}^{k_2}(M_2-j)}
\left(1+\frac{h}{g}\sum_{l=1}^{k_1+k_2}\frac{\frac{M}{2}-l}{z_{12}-1}\right)\el
&\quad\times
\sum_{m=0}^{k_1}\left[\left(-\frac{h}{g}\frac{1}{z_{12}-1}\right)^{k_1+n-2m} \,
z_{12}^{n-m}
{\scriptscriptstyle \left(\begin{array}{c} k_1 \\ m\end{array}\right)
\left(\begin{array}{c} k_2 \\ n-m\end{array}\right)}
\right.\el
&\qquad\times
\Biggl(1+\frac{h}{g}\sum_{p=0}^{m-1}\frac{z_{12}\left(\frac{M_2}{2}
+p\right)-\frac{M_1}{2}}{z_{12}-1}
+\frac{h}{g}\sum_{p=-m}^{k_2-n-1}\frac{z_{12}\left(\frac{M_1}{2}+p\right)-\frac{
M_2}{2}}{z_{12}-1}\el
&\qquad\quad\left.
+\frac{h}{2g}\bigl( k_2(n-m)-k_1 m-k_2^2\bigr)\Biggr)
\prod_{p=1+m}^{k_1}(M_1-p)\prod_{p=1}^{n-m}(M_2-K+n-p)\right]\!,
\label{Xcl}
\end{align}
where $\mathscr{D}_{cl}$ is $\mathcal{O}(g^{-1})$ term of $\mathscr{D}$ in
\eqref{coefX}.
Since the binomial coefficients force the index $m$ to be $m\leq \min\{k_1,n\}$,
we will discuss the two possible cases separately. They are the $n\neq k_1$ case
(off-diagonal sector) and the $n=k_1$ case (diagonal sector). \\


\noindent {\it Off-diagonal sector.} 
In the case when $n$ is different from $k_1$, it is further classified by two
more cases -- if $n$ is bigger or smaller than $k_1$.  
Firstly, in the $n>k_1$ case, the leading order of \eqref{Xcl} is
$\mathcal{O}(g^{-(n-k_1)})$ with $m=k_1$. Therefore the $\mathcal{O}(g^{-1})$ 
term, which contributes to the classical r-matrix,  is obtained by setting $n=k_1+1$. 
In this situation, the classical limit of \eqref{Xcl} turns out to be of a
simple form,
\begin{align}\label{offdiagr1}
\mathscr{X}^{k_1,k_2}_{k_1+1}\sim -\frac{h}{g} \frac{z_1}{z_1-z_2} k_2
(M_1-k_1-1)\;.
\end{align}
Secondary, in the $n<k_1$ case, the leading order is $\mathcal{O}(g^{-(k_1-n)})$
with $m=n$. Therefore, the $\mathcal{O}(g^{-1})$ contribution is given by $n=k_1-1$. 
In this case the amplitude becomes 
\begin{align}\label{offdiagr2}
\mathscr{X}^{k_1,k_2}_{k_1-1}\sim -\frac{h}{g} \frac{z_2}{z_1-z_2} k_1
(M_2-k_2-1)\;.
\end{align}
The other matrix elements do not contribute to the classical r-matrix.\\  


\noindent {\it Diagonal sector.} 
This is the $n=k_1$ case and it needs a more elaborate treatment in comparison
with the off-diagonal sector.
In this case the leading order in \eqref{Xcl} is $\mathcal{O}(1)$ with
$m=k_1=n$. Thus the classical limit turns out to be
\begin{align}
\mathscr{X}^{k_1,k_2}_{k_1}
&\sim 1+\mathscr{D}_{cl} -\frac{h}{2g}(k_1^2+k_2^2)
+\frac{h}{g}\frac{1}{z_1-z_2}\left[
\sum_{l=1}^{k_1+k_2}z_2\left(\frac{M}{2}-l\right)\right.\el
&\qquad\left.
+\sum_{p=0}^{k_1-1}\left(\frac{z_1M_2-z_2M_1}{2}+z_1p\right)
+\sum_{p=-k_1}^{k_2-k_1-1}\left(\frac{z_1M_1-z_2M_2}{2}+z_1p\right)\right]
.\label{diagr}
\end{align}
%


\paragraph{Full Rational Limit.}
It is noted that the classical limit still depends on the deformation parameter
$h$. This allows us to take $h\to0$ limit further, which corresponds to the case
``R(full)'' in the analysis of \cite{Bcl}. 
In this limit, the classical evaluation parameter \eqref{clev} reads, 
\begin{align}
z\sim 1-\frac{h}{g} u +\mathcal{O}(h^2),
\qquad\text{with}\qquad
u=x+\frac{1}{x}.
\end{align}
Then the off-diagonal elements of the classical r-matrix 
\eqref{offdiagr1} and \eqref{offdiagr2} turns out to be 
\begin{align}
\label{rfull1}
\mathscr{X}^{k_1,k_2}_{k_1+1}\sim \frac{1}{\delta u} k_2 (M_1-k_1-1)\;,
\qquad 
\mathscr{X}^{k_1,k_2}_{k_1-1}\sim \frac{1}{\delta u} k_1 (M_2-k_2-1)\;.
\end{align}
On the other hand, the diagonal elements \eqref{diagr} reduce to 
\begin{align}
\label{rfull2}
\mathscr{X}^{k_1,k_2}_{k_1}
&\sim 1+\mathscr{D}_{cl} 
-\frac{1}{\delta u}\left[\sum_{l=1}^{k_1+k_2}\left(\tfrac{M}{2}-l\right)
+\sum_{p=0}^{k_1-1}\left(-\tfrac{\delta M}{2}+p\right)
+\sum_{p=-k_1}^{k_2-k_1-1}\left(\tfrac{\delta M}{2}+p\right)\right].
\end{align}
The above expressions \eqref{rfull1} and \eqref{rfull2} agree with 
the classical limits of rational case \cite{ALT}.


\subsection{Scattering in subspace II}\label{sII}

The S-matrix in the subspace II is defined to be
\begin{align}
\S \,\stateB{k_1,k_2}_i &= \sum^{K}_{n=0} \sum^{4}_{j=1} |n,K\!-n\rangle^{\rm
II}_j \,\bigl(\mathscr{Y}^{k_1,k_2}_{n}\bigr)^j_i\,,
\end{align}
and the standard $4N+2$--dimensional basis is
\begin{align}
\stateB{k_1,k_2}_1 &= |0,1,k_1,M_1-k_1-1\rangle \otimes
|0,0,k_2,M_2-k_2\rangle,\el
\stateB{k_1,k_2}_2 &= |0,0,k_1,M_1-k_1\rangle \otimes
|0,1,k_2,M_2-k_2-1\rangle,\el
\stateB{k_1,k_2}_3 &= |0,1,k_1,M_1-k_1-1\rangle \otimes
|1,1,k_2-1,M_2-k_2-1\rangle,\el
\stateB{k_1,k_2}_4 &= |1,1,k_1-1,M_1-k_1-1\rangle \otimes
|0,1,k_2,M_2-k_2-1\rangle.
\end{align}
We shall express the coefficients $(\mathscr{Y}^{k_1,k_2}_{n})^j_i$ in terms of
already known $\mathscr{X}^{k_1,k_2}_{n}$ with the help of the charges $\DE_{2}$
and $\DE_{4}$ that relate the states in the subspace II to the states in
subspace I:
\begin{align}
\DE_2 \,\stateB{k_1,k_2}_j = Q_j(k_1,k_2) \, \stateA{k_1,k_2},\qquad
\DE_4 \,\stateB{k_1,k_2}_j = \widetilde{Q}_j(k_1,k_2) \,
\stateA{k_1,k_2}.\label{Qs}
\end{align}
The coefficients $Q_j(k_1,k_2)$, $\widetilde Q_j(k_1,k_2)$ and their partners
for $\DOE_{2}$ and $\DOE_{4}$ are spelled out in the Appendix \ref{app:A1}.

The strategy of finding $\mathscr{X}^{k_1,k_2}_{n}$ is the following. We start
by considering the matrix element
\begin{align}
\costateA{n,K\!-n}|\,\DOE_2\,\S\,\stateB{k_1,k_2}_i
&= \sum_{j=1}^4\sum_{m=0}^{K} \ \costateA{n,K\!-n} |\,\DOE_2\,\stateB{m,K\!-m}_j
\ \bigl(\mathscr{Y}^{k_1,k_2}_{m}\bigr)^j_i \el
&= \sum_{j=1}^4\sum_{m=0}^{K} \ \costateA{n,K\!-n}\stateA{m,K\!-m}\
Q^{op}_j(m,K\!-m)\ \bigl(\mathscr{Y}^{k_1,k_2}_{m}\bigr)^j_i  \el
&= \sum_{j=1}^4Q^{op}_j(n,K\!-n) \
\bigl(\mathscr{Y}^{k_1,k_2}_{n}\bigr)^j_i\,.\label{S2eq1}
\end{align}
Next, using the invariance of the S-matrix $\DOE_2\,\S=\S\,\DE_2$, we
rewrite \eqref{S2eq1} as
\begin{align}
\costateA{n,K\!-n}|\,\S\,\DE_2\,\stateB{k_1,k_2}_i 
&= \costateA{n,K\!-n}|\,\S\,\stateA{k_1,k_2} \,Q_i(k_1,k_2) \el 
&= \sum_{m=0}^{N}\ \costateA{n,K\!-n}\stateA{m,K\!-m}
\,\mathscr{X}^{k_1,k_2}_{m} \,Q_i(k_1,k_2) \el
&= \ \mathscr{X}^{k_1,k_2}_{n}\,Q_i(k_1,k_2).\label{S2eq2}
\end{align}
%
Likewise we get a similar set of relations by considering the
charge $E_4$. These relations can be conveniently summarized in terms of
a matrix equation
\begin{align}
&\begin{pmatrix}
  Q^{op}_1(n,K\!-n) & Q^{op}_2(n,K\!-n) & Q^{op}_3(n,K\!-n)  & Q^{op}_4(n,K\!-n)
\\
  \tilde{Q}^{op}_1(n,K\!-n)  & \tilde{Q}^{op}_2(n,K\!-n) &
\tilde{Q}^{op}_3(n,K\!-n) & \tilde{Q}^{op}_4(n,K\!-n)
\end{pmatrix}
\mathscr{Y}^{k_1,k_2}_{n} = \el
&\qquad\qquad\qquad\qquad\qquad =\mathscr{X}^{k_1,k_2}_n
\begin{pmatrix}
  Q_1(k_1,k_2) & Q_2(k_1,k_2) & Q_3(k_1,k_2)  & Q_4(k_1,k_2) \\
  \tilde{Q}_1(k_1,k_2)  & \tilde{Q}_2(k_1,k_2) & \tilde{Q}_3(k_1,k_2) &
\tilde{Q}_4(k_1,k_2)
\end{pmatrix},
\end{align}
giving a total number of 8 constraints. However, there is a further need of 8
more constraints. These can be obtained by considering a composite operator
\begin{align}
\check{E}_2 = e_0 \Bigl(e_1\,\hat F_1 F_3 F_2 + e_2\,F_1F_3F_{2} + e_3\,F_3F_{2}
F_1\Bigr),
\end{align}
where
\begin{align}
e_0 &= q^{1+K+\frac{M_1}{2}}(q^M z_1- q^{2K+2}z_2)^{-1},
 & e_1 &= (q-q^{-1}), \el
e_2 &= q^{M_2+2n}(q^{-2-2K}z_1- q^{2-M}z_2),
 & e_3 &= -q^{M_2+2n}(q^{-1-2K}z_1- q^{1-M}z_2),\label{E2coefs}
\end{align}
and its affine partner $\check{E}_4$. These operators act on the states in the
subspace II as
\begin{align}
\Delta\check{E}_2\stateB{k_1,k_2}_i =& Z_i(k_1,k_2)\stateA{k_1,k_2} +
Z^+_i(k_1,k_2)\stateA{k_1+1,k_2-1}\el
&\qquad\qquad\qquad\qquad + Z^-_i(k_1,k_2)\stateA{k_1-1,k_2+1},
\end{align}
giving
\begin{align}
&\costateA{n,K\!-n}|~\Delta^{\!op}\check{E}_2~\S~\stateB{k_1,k_2}_i 
=\sum_{j=1}^4\sum_{m=0}^{K} \
\costateA{n,K\!-n} |~\Delta^{\!op}\check{E}_2~\stateB{m,K-m}_j \
\bigl(\mathscr{Y}^{k_1,k_2}_{m}\bigr)^j_i\el
&\quad =\sum_{j=1}^4 \biggl(Z^{op}_j(n,K\!-n) \
\bigl(\mathscr{Y}^{k_1,k_2}_{n}\bigr)^j_i 
+ Z^{+,op}_j(n-1,K\!-n+1) \ \bigl(\mathscr{Y}^{k_1,k_2}_{n-1}\bigr)^j_i \el
&\qquad\qquad\qquad\qquad\qquad\qquad\qquad\quad\;\, 
+Z^{-,op}_j(n+1,K\!-n-1) \ \bigl(\mathscr{Y}^{k_1,k_2}_{n+1}\bigr)^j_i\biggr)
\,.\label{newE2rel}
\end{align}
The coefficients (\ref{E2coefs}) are chosen in a such way that the
`non-diagonal' part of this relation is vanishing, $Z^{+,op}_j(n-1,K\!-n+1)=
Z^{-,op}_j(n+1,K\!-n-1)=0$. Therefore the only surviving part of
(\ref{newE2rel}) is
\begin{align}
&\costateA{n,K\!-n}|~\Delta^{\!op}\check{E}_2~\S~\stateB{k_1,k_2}_i = 
\sum_{j=1}^4 Z^{op}_j(n,K\!-n) \ \bigl(\mathscr{Y}^{k_1,k_2}_{n}\bigr)^j_i\,.
\end{align}
This results in the following matrix equation for $Z^{op}_j(n,K\!-n)$:
\begin{align}
&\begin{pmatrix} Z^{op}_1(n,K\!-n) & Z^{op}_2(n,K\!-n) & Z^{op}_3(n,K\!-n) &
Z^{op}_4(n,K\!-n) \end{pmatrix} \mathscr{Y}^{k_1,k_2}_n  \el
&\quad =\begin{pmatrix} Z_1(k_1,k_2) & Z_2(k_1,k_2) & Z_3(k_1,k_2) &
Z_4(k_1,k_2) \end{pmatrix} \mathscr{X}^{k_1,k_2}_n \el
&\qquad + \begin{pmatrix} Z^+_1(k_1,k_2) & 0 & Z^-_3(k_1,k_2) & 0 \end{pmatrix}
\mathscr{X}^{k_1+1,k_2-1}_{n}\el  
&\qquad + \begin{pmatrix}0 & Z^-_2(k_1,k_2) & 0 & Z^-_4(k_1,k_2) \end{pmatrix}
\mathscr{X}^{k_1-1,k_2+1}_{n},\label{YX}
\end{align}
plus a similar set of equations arising from the affine charge $\check E_4$.
Both sets can further be united into a compact matrix form 
\begin{align}\label{eqY}
A\,\mathscr{Y}^{k_1,k_2}_n = B \mathscr{X}^{k_1,k_2}_n +
B^+\mathscr{X}^{k_1+1,k_2-1}_{n} + B^- \mathscr{X}^{k_1-1,k_2+1}_{n},
\end{align}
which multiplied from the left by $A^{-1}$ defines all coefficients of
$\mathscr{Y}^{k_1,k_2}_n$ in terms of already known $\mathscr{X}^{k_1,k_2}_{n}$,
$\mathscr{X}^{k_1\pm1,k_2\mp1}_{n}$. 
The explicit expressions of matrices $A$, $A^{-1}$, $B$, $B^{\pm}$, 
their $q\to1$ limit and the coefficients $Z_i(k_1,k_2)$, $Z^{op}_j(n,K\!-n)$ and
their affine partners are spelled out the Appendix \ref{app:A1}.  

To finalize we want to note that not all of the constraints in \eqref{YX} are
linearly independent. The set of independent constraints is chosen in such way
that the inverse matrix $A^{-1}$ would exist.


\subsection{Scattering in subspace III}\label{sIII}

We will compute the S-matrix components in the subspace III in a very similar
way as we did in the previous section for the scattering in subspace II. We start by
defining the S-matrix for the subspace III as
\begin{align}
\S \,\stateC{k_1,k_2}_i &= \sum^{K}_{n=0} \sum^{6}_{j=1} \stateC{n,K-n}_j 
\bigl(\mathscr{Z}^{k_1,k_2}_{n}\bigr)^j_i
\;,
\end{align}
where the standard basis for the $6N$-dimensional vector space is
\begin{align}
\stateC{k_1,k_2}_1 &= |0,0,k_1,M_1-k_1\rangle \otimes
|0,0,k_2,M_2-k_2\rangle,\el
\stateC{k_1,k_2}_2 &= |0,0,k_1,M_1-k_1\rangle \otimes
|1,1,k_2-1,M_2-k_2-1\rangle,\el
\stateC{k_1,k_2}_3 &= |1,1,k_1-1,M_1-k_1-1\rangle \otimes
|0,0,k_2,M_2-k_2\rangle,\el
\stateC{k_1,k_2}_4 &= |1,1,k_1-1,M_1-k_1-1\rangle \otimes
|1,1,k_2-1,M_2-k_2-1\rangle,\el
\stateC{k_1,k_2}_5 &= |1,0,k_1-1,M_1-k_1\rangle \otimes
|0,1,k_2,M_2-k_2-1\rangle,\el
\stateC{k_1,k_2}_6 &= |0,1,k_1,M_1-k_1-1\rangle \otimes
|1,0,k_2-1,M_2-k_2\rangle.
\end{align}

Next we shall employ the same strategy as before. 
We perform the same steps as in (\ref{S2eq1}) and (\ref{S2eq2}) only with
$\DOE_2$, giving
\begin{align}
_i\costateB{n,K\!-n}|\,\DOE_2\,\S\,\stateC{k_1,k_2}_j 
&=\sum_{l=1}^6\bigl(G^{op}(n,K-n)\bigr)^i_l\bigl(\mathscr{Z}^{k_1,k_2}_{n}
\bigr)^l_j,\el
_i\costateB{n,K\!-n}|\,\S\,\DE_2\,\stateC{k_1,k_2}_j 
&=\sum_{m=1}^4\bigl(\mathscr{Y}^{k_1,k_2}_{n}\bigr)^i_m
\bigl(G(k_1,k_2)\bigr)^m_j,
\end{align}
where $G^{(op)}$ are the matrix representations of the charges $\Delta^{\!(op)}E_2$.
Once again these equations (together with the affine ones coming from $E_4$) 
do not provide enough constraints to define the matrix 
$\mathscr{Z}^{k_1,k_2}_{n}$ uniquely, and we need additional constraints. 
They are obtained with the help of $\Delta^{\! (op)}(F_3 F_2)$, namely
\begin{align}
_i\costateB{n-\theta_i,K\!-n+\theta_i-1}|\,{\rm \Delta}^{\!
op}(F_3F_2)\,\S\,\stateC{k_1,k_2}_j 
&=\sum_{l=1}^6\bigl(H^{op}(n,n-K)\bigr)^i_l
\bigl(\mathscr{Z}^{k_1,k_2}_{n}\bigr)^l_j\,,
\el
_i\costateB{n-\theta_i,K\!-n+\theta_i-1}|\,\S\,{\rm
\Delta}(F_3F_2)\,\stateC{k_1,k_2}_j 
&=\sum_{m=1}^4 \bigl(\overline{\mathscr{Y}}^{k_1,k_2}_{n}\bigr)^i_m
\bigl(H(k_1,k_2)\bigr)^m_j\,,
\end{align}
where $\theta_i$ is defined by $\theta_i=(1-(-1)^i)/2$ 
and $H^{(op)}$ is the matrix representation of ${\rm \Delta}^{\! (op)}(F_3F_2)$.
Here we have also introduced $\overline{\mathscr{Y}}^{k_1,k_2}_{n}$ as 
\begin{align}
\bigl(\overline{\mathscr{Y}}^{k_1,k_2}_{n}\bigr)^i_j
=\bigl(\mathscr{Y}^{k_1-\theta_j,k_2+\theta_j-1}_{n-\theta_i}\bigr)^i_j\,.
\end{align}
These equations may be written in a compact way using matrix notation
\begin{align}
G^{op}(n,K\!-n)\,\mathscr{Z}^{k_1,k_2}_{n} &=
\mathscr{Y}^{k_1,k_2}_{n}\,G(k_1,k_2), \el
H^{op}(n,K-n)\,\mathscr{Z}^{k_1,k_2}_{n} &=
\overline{\mathscr{Y}}^{k_1,k_2}_{n}\,H(k_1,k_2).
\label{S3eq2}
\end{align}
The explicit realization of the matrices in the expressions above are spelled
out in the Appendix \ref{app:A2}. 

Similarly as in the previous case, not all rows and columns of $G^{(op)}$ and
$H^{(op)}$ are linearly independent, thus we have to select the independent ones only. 
Therefore by taking the following linear combinations,
\begin{align}
\overline{G}^{(op)}=q^{K-n-\frac{M_2}{2}} \bigl(\tl a_2G^{(op)} - a_2\wtl
G^{(op)}\bigr)
\quad\; \text{and}\quad\; 
\overline{H}^{(op)}=\tl c_2V_1H^{(op)} - c_2V_1^{-1}\wtl H^{(op)}\,,  
\end{align}
where the tilded matrices are the affine counterparts and selecting the first
three rows of each, we are able to combine them into the non-singular quadratic matrix $A$
($6\times6$) and the rectangular matrix $B$ ($8\times6$) as follows ($j=1,\cdots, 6$),  
\begin{align}
(A)^i_j=\begin{cases} (\overline{G}^{op})^i_j\,, & i=1,2,3\,, \\
(\overline{H}^{op})^{i-3}_j\,, & i=4,5,6\,, \end{cases}
\qquad \text{and}\qquad 
(B)^i_j=\begin{cases} (\overline{G})^i_j\,, & i=1,2,3,4\,, \\
(\overline{H})^{i-4}_j\,, & i=5,6,7,8\;. \end{cases}
\end{align}
This approach let us to rewrite the constraints \eqref{S3eq2} in terms of a single matrix relation
\begin{align}\label{eqZ}
A\,\mathscr{Z}^{k_1,k_2}_{n} = \check{\mathscr{Y}}^{k_1,k_2}_n B 
\qquad\text{giving}\qquad 
\mathscr{Z}^{k_1,k_2}_{n} = A^{-1}\,\check{\mathscr{Y}}^{k_1,k_2}_n B\,.
\end{align}
This relation let us to obtain any matrix element
$(\mathscr{Z}^{k_1,k_2}_{n})^i_j$ of the scattering in the subspace III. 
Here we have also introduced the block diagonal matrix 
$\check{\mathscr{Y}}^{k,l}_n$ ($6\times 8$) as  
\begin{align}
\bigl(\check{\mathscr{Y}}^{k,l}_n\bigr)^i_j\,,
=\begin{cases} \bigl({\mathscr{Y}}^{k,l}_n\bigr)^i_j\,, & i=1,2,3\,,
\quad\text{and}\quad j=1,2,3,4\,, \\ 
\bigl(\overline{\mathscr{Y}}^{k,l}_n\bigr)^{i-3}_{j-4}\,, & i=4,5,6\,,
\quad\text{and}\quad j=5,6,7,8\,, \\ 
0\,, & \text{the rest}\,. \end{cases}
\end{align}
The explicit form of matrices $A$, $A^{-1}$, $B$ and their $q\to1$ limit are
given in Appendix \ref{app:A2}.


\section{Special cases of the S-matrix}\label{sec:SpecialCases}

In this section we consider the reduction of the S-matrix in the case when one
or both factors of the tensor space \eqref{bstates} are transforming in the
fundamental representation.


\subsection{Fundamental S-matrix}

As a most simple case of the derivations presented in section \ref{sec:Smat}, we
want to compute the fundamental S-matrix found in \cite{BK}. The fundamental
representation is defined by setting $M_1=M_2=1$ and the corresponding S-matrix
is $16\times16$ -- dimensional. In order to make the comparison with \cite{BK}
more explicit, let us denote
\begin{align}
&\ad_{1,2} = \phi^{1,2}, \qquad\qquad \text{and} \qquad\qquad \ad_{3,4} =
\psi^{1,2}.
\end{align}
Then, starting with the subspaces I and Ib, we find
\begin{align}
\S\,|\psi^{\alpha}\psi^{\alpha}\rangle =
\mathscr{D}\,|\psi^{\alpha}\psi^{\alpha}\rangle,
\end{align}
where $\mathscr{D}$ is given by (\ref{eqn;D}). Further, due to our normalization
\begin{align}
\S\,|\phi^{a}\phi^{a}\rangle = |\phi^{a}\phi^{a}\rangle.
\end{align}
Here we would like to remark that our normalization differs from \cite{BK} where
the S-matrix is normalized such that $\S\,|\psi^{\alpha}\psi^{\alpha}\rangle =
-|\psi^{\alpha}\psi^{\alpha}\rangle$. In other words, the quantities given here
need to be divided by an additional factor of $\mathscr{D}$.

Next we proceed to the subspaces II and IIb. For the subspace II (and
analogously for IIb) the parameters $k_1,\,k_2,\,n$ indexing the matrix
$\mathscr{Y}$ can take the values 0 and 1, but fortunately, we find that
$\mathscr{Y}$ is the same for both of these values. Next it is easy to observe
that the matrices $A$ \eqref{sIIA} and $B$ \eqref{sIIB} get reduced to the upper
left $2\times2$ blocks
\begin{align}
A = \begin{pmatrix}
   - a_2  & \;q^{1/2} U_2V_2 a_1\\
   -\tl{a}_2 & \;q^{1/2} \tU_2\tV_2 \tl{a}_1
    \end{pmatrix},
&&
B = \begin{pmatrix}
   - a_2 \sqrt{q} U_1 V_1 & a_1\\
   -\tl{a}_2 \sqrt{q} \tU_1 \tV_1 & \tl{a}_1
    \end{pmatrix},
\end{align}
while the matrices $B^+$ and $B^-$ do not contribute at all. This gives the
following solution of \eqref{eqY}
\begin{align}
\mathscr{Y}^{0,0}_0 &= \mathscr{D} \left(
\begin{array}{cc}
 \frac{\sqrt{q}(a_2 \tl{a}_1 U_1^2 V_1^2-a_1 \tl{a}_2 U_2^2 V_2^2)}{U_1 V_1(a_2
\tl{a}_1-a_1 \tl{a}_2 U_2^2 V_2^2)} & \frac{a_1 \tl{a}_1 (1-U_2^2 V_2^2)}{a_1
\tl{a}_2 U_2^2 V_2^2-a_2 \tl{a}_1} \\
 \frac{a_2 \tl{a}_2 U_2 (U_1^2 V_1^2-1) V_2}{U_1 V_1(a_2 \tl{a}_1-a_1 \tl{a}_2
U_2^2 V_2^2)} & 
 \frac{(a_2 \tl{a}_1-a_1 \tl{a}_2) U_2 V_2}{\sqrt{q} (a_2 \tl{a}_1-a_1 \tl{a}_2
U_2^2 V_2^2)}
\end{array}
\right)\el
 &= \left(
\begin{array}{cc}
\scriptstyle{q^{1/2}U_2V_2}\frac{x^-_2 - x^-_1}{x^+_2 - x^-_1} & 
\frac{\gamma_1}{\gamma_2}\frac{U_2V_2}{U_1V_1}\,\frac{x^+_2 - x^-_2}{x^+_2 -
x^-_1} 
\\
\frac{\gamma_2}{\gamma_1}\, \frac{x^+_1 - x^-_1}{x^+_2 - x^-_1}& 
\frac{1}{q^{1/2} U_1 V_1}\, \frac{x^+_2-x^+_1}{x^+_2-x^-_1}
\end{array}
\right).
\end{align}
Then the corresponding explicit form of the fundamental S-matrix acting on the
inequivalent states is
\begin{align}
&\S|\psi^\alpha\phi^b\rangle 
=q^{1/2}U_2V_2\frac{x^-_2 - x^-_1}{x^+_2 - x^-_1}|\psi^\alpha\phi^b\rangle 
+\frac{\gamma_2}{\gamma_1}\, \frac{x^+_1 - x^-_1}{x^+_2 -
x^-_1}|\phi^b\psi^\alpha\rangle\,,\el
&\S|\phi^a\psi^\beta\rangle 
=\frac{\gamma_1}{\gamma_2}\frac{U_2V_2}{U_1V_1}\,\frac{x^+_2 - x^-_2}{x^+_2 -
x^-_1}|\psi^\beta\phi^a\rangle 
+\frac{1}{q^{1/2} U_1 V_1}\,
\frac{x^+_2-x^+_1}{x^+_2-x^-_1}|\phi^a\psi^\beta\rangle\,.
\label{sII:fund}
\end{align}

Finally we turn to the subspace III which is four dimensional in this case.
Analogously to our strategy presented section \ref{sII}, we inspect the action
of $\DE_{2}$ and $\DE_4$ obtaining
\begin{align}
&\DE_2 \stateC{1,0}_1 = \frac{U_1V_1}{\sqrt{q}}a_2 \stateB{1,0}_2, && 
\DE_2 \stateC{1,0}_5 = b_1 \stateB{1,0}_2, \nonumber\\
&\DE_2 \stateC{0,1}_1 = a_1 \stateB{0,0}_1, && 
\DE_2 \stateC{0,1}_6 =-U_1V_1\sqrt{q}\, b_2 \stateB{0,0}_1,
\end{align}
plus similar expressions for $E_4$. For completeness, let us spell out the
opposite coproduct as well
\begin{align}
&\DOE_2 \stateC{1,0}_1 = a_2 \stateB{1,0}_2, && 
\DOE_2 \stateC{1,0}_5 = b_1 U_2V_2\sqrt{q} \stateB{1,0}_2, \nonumber\\
&\DOE_2 \stateC{0,1}_1 = a_1 \frac{U_2V_2}{\sqrt{q}}\stateB{0,0}_1, && 
\DOE_2 \stateC{0,1}_6 = -b_2 \stateB{0,0}_1.
\end{align}
The equation \eqref{eqZ} in this case becomes
\begin{align}
\begin{pmatrix}
a_2 & b_1 \sqrt{q} U_2V_2 \\
\tl{a}_2 & \tl{b}_1 \sqrt{q} \tU_2\tV_2
\end{pmatrix}
\begin{pmatrix}
(\mathscr{Z}^{1,0}_{1})^1_1 & (\mathscr{Z}^{1,0}_{1})^5_1\\
(\mathscr{Z}^{1,0}_{1})^5_1 & (\mathscr{Z}^{1,0}_{1})^5_5
\end{pmatrix}=
\begin{pmatrix}
\frac{U_1V_1}{\sqrt{q}}a_2 & b_1 \\
\frac{\tU_1\tV_1}{\sqrt{q}}\tl{a}_2 & \tl{b}_1
\end{pmatrix} (\mathscr{Y}^{1,0}_{1})^2_2\,,
\end{align}
the explicit solution of which is
\begin{align}
\begin{pmatrix}
(\mathscr{Z}^{1,0}_{1})^1_1 & (\mathscr{Z}^{1,0}_{1})^5_1\\
(\mathscr{Z}^{1,0}_{1})^5_1 & (\mathscr{Z}^{1,0}_{1})^5_5
\end{pmatrix}=
\left(
\begin{array}{cc}
 \frac{(1-x^-_2 x^+_1) (x^+_1-x^+_2)}{(1-x^-_1x^-_2) (x^-_1-x^+_2)}
\frac{x^-_1}{q x^+_1}& 
\frac{\alpha (x^-_1-x^+_1) (x_2^- - x^+_2) (x^+_1 - x^+_2)}{\sqrt{q} U_1 V_1
\gamma_1\gamma_2(x^-_1x^-_2-1) 
(x^-_1 - x^+_2)} \\
\frac{\gamma_1\gamma_2(x^+_1 - x^+_2)}{U_2V_2 \alpha 
(1-x^-_1x^-_2)(x^+_2-x^-_1)}\frac{x^-_1}{q^{3/2} x^+_1}&
\frac{(1-x^-_1x^+_2) (x^+_1 - x^+_2)}{(1-x^-_1x^-_2)(x^-_1 -
x^+_2)}\frac{U_2V_2}{U_1V_1}\frac{x^-_2}{q x^+_2}
\end{array}
\right).
\end{align}
The remaining matrix elements are then easily deduced from similar derivations.
These results are in agreement with \cite{BK}. For a complete list of all the
scattering elements we refer to the Appendix \ref{app:fundS}.


\subsection{The S-matrix S\texorpdfstring{$_{Q1}$}{SQ1}}

In this section we will derive the S-matrix describing the scattering of an
arbitrary bound state with a fundamental one, $\S_{Q1}$. Once again, we will
follow the derivations performed in section \ref{sec:Smat} step by step. 
First, by setting $M_2=1$, we find that the states in subspaces I and Ib scatter
almost trivially
\begin{align}
\S\,\stateA{k,0} = \mathscr{D}\,\stateA{k,0}.
\end{align}
However the scattering in the subspace II does not get simplified that much.
Nevertheless, for fixed $k_1+k_2$, the corresponding vector space gets
restricted to
\begin{align}\label{SQ1:sIIIstates}
\{
\stateB{k_1,0}_1,\,\stateB{k_1-1,1}_1,\,\stateB{k_1,0}_2,\,\stateB{k_1,0}_4\}.
\end{align}
This is because the states $\stateB{k_1,k_2}_3$ have $M_2\geq 2$ and thus they
are not present. 
By reducing our general expressions to accommodate these 4 states, we are lead
to 16 inequivalent scattering elements, however we found 2 of them to be
vanishing. The rest may be casted in quite compact form as
\begin{align}
\S\,\stateB{k,0}_1 &= (\mathscr{Y}^{k,0}_{0})^1_1\stateB{k,0}_1 +
(\mathscr{Y}^{k,0}_{1})^1_1\stateB{k\!-\!1,1}_1 +
(\mathscr{Y}^{k,0}_{0})^2_1\stateB{k,0}_2 +
(\mathscr{Y}^{k,0}_{0})^4_1\stateB{k,0}_4 \,,\el
\S\,\stateB{k\!-\!1,1}_1 &= (\mathscr{Y}^{k\!-\!1,1}_{0})^1_1\stateB{k,0}_1 +
(\mathscr{Y}^{k\!-\!1,1}_{1})^1_1\stateB{k\!-\!1,1}_1 +
(\mathscr{Y}^{k\!-\!1,1}_{0})^2_1\stateB{k,0}_2 +
(\mathscr{Y}^{k\!-\!1,1}_{0})^4_1\stateB{k,0}_4 \,,\el
\S\,\stateB{k,0}_2 &= (\mathscr{Y}^{k,0}_{0})^1_2\stateB{k,0}_1 +
(\mathscr{Y}^{k,0}_{1})^1_2\stateB{k\!-\!1,1}_1 +
(\mathscr{Y}^{k,0}_{0})^2_2\stateB{k,0}_2\,,\el
\S\,\stateB{k,0}_4 &= (\mathscr{Y}^{k,0}_{0})^1_4\stateB{k,0}_1 +
(\mathscr{Y}^{k,0}_{1})^1_4\stateB{k\!-\!1,1}_1 +
(\mathscr{Y}^{k,0}_{0})^4_4\stateB{k,0}_4\,.
\end{align}
The explicit expressions of the coefficients above are given in Appendix
\ref{app:SQ1}. Upon setting $M_1=1$ the coefficients with indices $1$ and $2$
reduce to the ones of the fundamental S-matrix \eqref{sII:fund} derived
previously.

The scattering in the subspace III simplifies considerably. It is easy to see,
that the states $\stateC{k_1,k_2}_{2,4}$ need not to be considered. Thus we are
led to the reduced case of our general expressions for subspace III that involve
the states \eqref{SQ1:sIIIstates} and
\begin{align}
\{ \stateC{k,0}_1, \,\stateC{k,0}_3, \,\stateC{k,0}_5, \,\stateC{k-1,1}_1,
\,\stateC{k-1,1}_3, \,\stateC{k-1,1}_6\}
\end{align}
only. However, there is a more straightforward way to obtain the S-matrix in
this particular case. 

There are 36 scattering coefficients in subspace III that need to be determined,
but not all of them are independent. Firstly we can relate the half of them to
the other half by considering the identity
\begin{align}
&\DE_3\stateA{k-1,0} =\stateC{k,0}_5 + q^{-1} \stateC{k-1,1}_6,
\end{align}
giving
\begin{align}\label{eqn;SQ1XtoZ}
\S\,\stateC{k-1,1}_6 =  \mathscr{D}\,\bigl( \stateC{k,0}_5 +q\,
\stateC{k-1,1}_6\bigr)  -q\, \S\,\stateC{k,0}_5 \,.
\end{align}
Subsequently we can express the states $\stateC{k-1,1}_1,\;\stateC{k-1,1}_3$ as
follows
\begin{align}\label{eqn;SQ1ZtoZ}
&\frac{\DF_1\DE_1-q[k]_q[M-k+1]_q}{[k]_q}\,\stateC{k,0}_1 =\stateC{k-1,1}_1,\el 
&\frac{\DF_1\DE_1-q[k-1]_q[M-k]_q}{[k-1]_q}\,\stateC{k,0}_3 =\stateC{k-1,1}_3.
\end{align}
The explicit constraints that follow from these identities are listed in the
Appendix \ref{app:SQ1}.

Then instead of reducing the general expression of the  matrix $\mathscr{Z}$, we
follow its derivation path. By considering the action of the charges $F_{2}$ and
$F_4$ on the subspace II states we are able to find simple expressions that
relate subspaces III to subspace II as
\begin{align}
&\stateC{k,0}_1 = \frac{\tilde{c}_1 V_2\DF_2 - c_1 \tV_2\DF_4}{\tilde{c}_1
d_2\tU_1V_2 - c_1\tilde{d}_2 U_1 \tV_2}\,\stateB{k,0}_2, &&
\stateC{k,0}_3= \frac{\tilde{d}_1V_2\DF_2 - d_1 \tV_2\DF_4}{\tl{d}_1d_2\tU_1V_2
- \tl{d}_2d_1U_1\tV_2}\,\stateB{k,0}_4\,,\el
&\stateC{k,0}_5 =\frac{\sqrt{q}}{[k]_q}
\frac{\tl{d}_2U_1\DF_2-d_2\tU_1\DF_4}{c_1\tl{d}_2U_1\tV_2-\tl{c}_1d_2\tU_1V_2}\,
\stateB{k,0}_2. 
\end{align}
This approach let us to find the expressions of the matrix elements of
$\mathscr{Z}$ in terms of the matrix elements of $\mathscr{Y}$ for this
particular case in quite an easy way. The explicit expressions are once again
given in the Appendix \ref{app:SQ1}.


\section{Discussion and outlook}

In this work we have constructed the supersymmetric short representations of the
quantum affine algebra $\afQ$ based on the centrally extended
$\alg{su}(2|2)$ algebra by making use of the quantum oscillator
algebra. These representations are of great importance as they
accommodate the bound states of the model. We found that the bound state
representations of the affine extension show a lot of similarities with
the fundamental one constructed in \cite{BGM}. In particular, we found that the
affine central elements are inverse to their non-affine partners, exactly as for
the fundamental representation. Moreover, the parameterization can be derived
from the fundamental one by the simple map $(q,g)\to(q^M,g/[M]_q)$.

The affine extension plays a key role in the construction of the bound state
S-matrix. Indeed, the affine generators $E_4$ and $F_4$ are crucial in
constructing the blocks $\mathscr{X}$ and $\mathscr{Y}$. In other words, the bound state
S-matrix is uniquely fixed up to the overall dressing phase by requiring
invariance under the affine algebra $\afQ$.

We have also spelled out the explicit coefficients of the S-matrix when one of
the spaces corresponds to the fundamental representation. And in particular, we have
checked that our formalism correctly reproduces the fundamental S-matrix found
in \cite{BK}. Furthermore, our results are in a very good agreement with those
of \cite{ALT}, where a similar derivation based on the Yangian symmetry related
to the same underlying Lie superalgebra was performed. More precisely, the
S-matrix we have obtained in the $q\to1$ limit for the subspace I reduces to the one found in \cite{ALT}. However we can not make a direct
comparison for subspaces II and III as the intermediate expressions are
different. This is due to the fact that affine rather than Yangian
generators are used. Nevertheless, the expressions we have obtained in this work
are of more symmetric form than those of \cite{ALT}. This is an expected result,
as the deformed quantum affine algebra itself is of more symmetric form than its
Yangian limit. 

We have not checked the Yang-Baxter equation in full generality due to this
being extremely challenging from the technical side.
However, we have performed a series of checks for a wide variety of states using
numerical computations and found that it was perfectly satisfied. A more detailed discussion on this point is given in 
Appendix \ref{sec;YBE}. 

In order to complete the investigations concerning the S-matrix it would be
interesting to consider the crossing symmetry and the corresponding solutions
for a $q$-deformed dressing phase, which at the moment are not known.

A particularly interesting direction for future research would be to
study the representations and their S-matrices for $q$ being a root of
unity. It is well known that the representation theory for these values of $q$
differs from the one for real $q$. Due to the bound state map being of the form
$q\to q^M$, it is not difficult to see that there appears to be some intrinsic
periodicity to these representations. One could hope, for example in the context
of the thermodynamic Bethe ansatz, that this would result in a finite number of
bound states, leading potentially to some useful insights in this area. 

A different topic related to this, would be to investigate the algebraic Bethe
ansatz and the bound state transfer matrices. This could perhaps be used to find
a $q$-deformed version of the $T$-system. 

One more possible direction of investigations is to consider the boundary
conditions and boundary scattering for the deformed Hubbard Chain. A good
starting point for this approach would be to consider the boundary conditions
equivalent to the ones of the $Y=0$ and $Z=0$ giant gravitons in the framework
of the AdS/CFT correspondence \cite{MN}. We expect some sort of deformed
(twisted) coideal subalgebra of $\afQ$ to be governing the boundary scattering
of the aforementioned type that in the rational limit would reproduce the
twisted Yangian algebras constructed in \cite{AN,MR1,MR2}.

Other open questions include
the search of the complete algebraic R-matrix and a
detailed investigation of its classical limit along the lines of
\cite{Bcl}. It would also be interesting to extend the classical limit to the
next order. For the undeformed case it was found that this order coincides with
the square of the classical $r$-matrix \cite{LeeuwThesis}. \\

\paragraph{Acknowledgments.} We would like to thank 
G.\ Arutyunov, N.\ Beisert, N.\ MacKay, S.\ Moriyama and A.\ Torrielli 
for helpful comments and discussions. 
T.M. also thanks Y.\ Kajihara, H.\ Konno, K.\ Oshima and H.\ Yamane for helpful
comments and valuable discussions.  
T.M. would like to warmly thank Max-Planck-Institut f\"ur Gravitationsphysik,
Albert-Einstein-Institut, in Potsdam for the hospitality where main part of 
this work was performed. V.R. also thanks the UK EPSRC for funding under 
grant EP/H000054/1.


\newpage

\appendix


\section{Elements of the S-matrix}

In this Appendix we have spelled out various coefficients and matrices that have
been heavily used in the intermediate steps in deriving the final expressions of
the S-matrix for the subspaces II and III.


\subsection{Subspace II}\label{app:A1}

The coefficients for the charge $\DE_2$ in \eqref{Qs} are
\begin{align}
Q_1(k_1,k_2) &= -q^{M_1/2-k_1}\,a_2 \,U_1 V_1 \,[\bar{k}_2+1]_q, && Q_2(k_1,k_2)
= a_1 \,[\bar{k}_1+1]_q,\el
Q_3(k_1,k_2) &= -q^{M_1/2-k_1}\,b_2 \,U_1 V_1, && Q_4(k_1,k_2) = b_1.
\end{align}
Similarly, the coefficients for the charge $\DOE_2$ are
\begin{align}
Q^{op}_1(k_1,k_2) &= -a_2 \,[\bar{k}_2+1]_q, && Q^{op}_2(k_1,k_2) =
q^{M_2/2-k_2} \,a_1 \,U_2 V_2\,[\bar{k}_1+1]_q,\el
Q^{op}_3(k_1,k_2) &= -b_2, && Q^{op}_4(k_1,k_2) = q^{M_2/2-k_2}\,b_1 \,U_2 V_2.
\end{align}
By replacing $a,b \to \tilde a,\tilde b$ and $U,V \to \tU,\tV$, 
one obtains $\widetilde Q_i(k_1,k_2)$ and $\widetilde Q^{op}_i(k_1,k_2)$ related
to the affine charge $E_4$. \\

\noindent The coefficients in \eqref{YX} are
\begin{align}
&Z^{op}_1(n,K\!-n) = c_2\, \tV_1\, [M_2-K+n]_q, && Z^{op}_2(n,K\!-n) = c_1\,
\tU_2\, [n-M_1]_q\, q^{n-K-\frac{M_1}{2}},\el
&Z^{op}_3(n,K\!-n) = d_2\, \tV_1\, q^{-M_2}, && Z^{op}_4(n,K\!-n) = -d_1\,
\tU_2\, q^{n-K+\frac{M_1}{2}}.
\end{align}
and 
\begin{align}
&Z_1(k_1,k_2) = \frac{c_2 \tU_1\,[\bar{k}_2+1]_q}{q^M z_{12} - q^{2(K+1)}}
 q^{M_1/2-k_1+M_2}\Bigl(q^{2n}z_{12}-q^{\delta
M}\bigl(q^{2(n-\bar{k}_1)}-1\bigr)-q^{2k_2+\delta M} \Bigr), \el
&Z_2(k_1,k_2) = \frac{z_{12}\, c_1 {\tV_2}\,[\bar{k}_1+1]_q}{q^M z_{12} -
q^{2(K+1)}}
q^{{-\delta M/2}+2}\Bigl(q^{2n}z_{21}-q^{\delta
M}\bigl(q^{2(n{+\bar{k}_2})}-q^{2K}\bigr)-q^{2k_2+\delta M} \Bigr), \el
 &Z_3(k_1,k_2) = \frac{d_2 \tU_1}{q^M z_{12} - q^{2(K+1)}}
 q^{M_1/2-k_1}\Bigl(q^{2n}z_{12}-q^{M}\bigl(q^{2(n-\bar{k}_1)}-1\bigr)-q^{
2k_2+\delta M} \Bigr), \el
&Z_4(k_1,k_2) = \frac{z_{12}\, d_1 \tV_2}{q^M z_{12} - q^{2(K+1)}}
q^{M/2+2}\Bigl(q^{2n}z_{21}-q^{-M}\bigl(q^{2(n{+\bar{k}_2})}-q^{2K}\bigr)-q^{
2k_2+\delta M} \Bigr).
\nonumber 
\end{align}
%

\noindent The matrices in \eqref{eqY} are defined as
\begin{align}
A = 
\begin{pmatrix}
Q^{op}_1(n,K\!-n) & Q^{op}_2(n,K\!-n) & Q^{op}_3(n,K\!-n) & Q^{op}_4(n,K\!-n) \\
\tilde{Q}^{op}_1(n,K\!-n) & \tilde{Q}^{op}_2(n,K\!-n) &
\tilde{Q}^{op}_3(n,K\!-n) & \tilde{Q}^{op}_4(n,K\!-n) \\
Z^{op}_1(n,K\!-n) & Z^{op}_2(n,K\!-n) & Z^{op}_3(n,K\!-n) & Z^{op}_4(n,K\!-n) \\
\tilde{Z}^{op}_1(n,K\!-n) & \tilde{Z}^{op}_2(n,K\!-n) &
\tilde{Z}^{op}_3(n,K\!-n) & \tilde{Z}^{op}_4(n,K\!-n)
\end{pmatrix},\label{sIIA}
\end{align}
\begin{align}
B = 
\begin{pmatrix}
Q_1(k_1,k_2) & Q_2(k_1,k_2) & Q_3(k_1,k_2) & Q_4(k_1,k_2) \\
\tilde{Q}_1(k_1,k_2) & \tilde{Q}_2(k_1,k_2) & \tilde{Q}_3(k_1,k_2) &
\tilde{Q}_4(k_1,k_2) \\
Z_1(k_1,k_2) & Z_2(k_1,k_2) & Z_3(k_1,k_2) & Z_4(k_1,k_2) \\
\tilde{Z}_1(k_1,k_2) & \tilde{Z}_2(k_1,k_2) & \tilde{Z}_3(k_1,k_2) &
\tilde{Z}_4(k_1,k_2)
\end{pmatrix},\label{sIIB}
\end{align}
and
\begin{align}
&B^+ = 
\begin{pmatrix}
0 & 0 & 0 & 0 \\
0 & 0 & 0 & 0 \\
Z^+_1(k_1,k_2) & 0 & Z^+_3(k_1,k_2) & 0 \\
\tilde{Z}^+_1(k_1,k_2) & 0 & \tilde{Z}^+_3(k_1,k_2) & 0
\end{pmatrix},
&&
B^- = 
\begin{pmatrix}
0 & 0 & 0 & 0 \\
0 & 0 & 0 & 0 \\
0 & Z^-_2(k_1,k_2) & 0 & Z^-_4(k_1,k_2) \\
0 & \tilde{Z}^-_2(k_1,k_2) & 0 & \tilde{Z}^-_4(k_1,k_2)
\end{pmatrix}.
\end{align}
The latter two have a quite compact explicit form
\begin{align}
&B^+ = 
[\bar{k}_1]_q\frac{q^{1+k_1-k_2-\frac{M_1}{2}}}{(q-q^{-1})^{-1}}
\frac{q^{M_1+2k_2}z_{12}-q^{{M_2+2(n+1)}}}{q^M z_{12} - q^{2(K+1)}} 
\begin{pmatrix}
0 & 0 & 0 & 0 \\
0 & 0 & 0 & 0 \\
-c_2\tU_1 [k_2]_q & 0 & d_2\tU_1 & 0 \\
-\tilde{c}_2 U_1 [k_2]_q & 0 & \tilde{d}_2 U_1 & 0
\end{pmatrix},\label{sIIBP} \\
&
B^- =
[\bar{k}_2]_q \frac{q^{1-k_1 +\frac{\delta M}{2}}}{(q-q^{-1})^{-1}}
\frac{q^{M_2+2n} z_{12}-q^{M_1+2(k_2+1)}}{q^M z_{12} - q^{2(K+1)}}  
\begin{pmatrix}
0 & 0 & 0 & 0 \\
0 & 0 & 0 & 0 \\
0 & -c_1\tV_2 [k_1]_q & 0 & d_1\tV_2 \\
0 & -\tilde{c}_1 V_2[k_1]_q & 0 & \tilde{d}_1 V_2
\end{pmatrix},\label{sIIBM}
\end{align}
%


\noindent The inverse of $A$ has a very complex form, however it can be
decomposed intro three quite compact matrices as $A^{-1} = CVD$, where
\begin{align}
&C = \left(
\begin{array}{cccc}
 \frac{z_{12} \tilde{b}_2}{[M_2-K+n]_q} & 0 & \frac{z_{12} \tilde{\alpha}
b_2}{[M_2-K+n]_q} & 0 \\
 0 & \frac{q^{K-\frac{M_2}{2}-n} \tilde{\alpha} b_1 U_2 V_2}{[n-M_1]_q} & 0 &
   \frac{q^{K-\frac{M_2}{2}-n} \tilde{b}_1}{[M_1-n]_qU_2V_2} \\
 -z_{12} \tilde{a}_2 & 0 & -z_{12} \tilde{\alpha}a_2 & 0 \\
 0 & q^{K-\frac{M_2}{2}-n} \tilde{\alpha} a_1 U_2 V_2 & 0 &
-\frac{q^{K-\frac{M_2}{2}-n}
   \tilde{a}_1}{U_2V_2}
\end{array}
\right),\\\el
& D = \mathrm{diag}\left(
\frac{i g \xi }{\tilde g \alpha \tilde{\alpha} z_2 },\,\frac{i g \xi }{\tilde g
\alpha \tilde{\alpha}^2 z_2},
\,\frac{q^{\frac{M_2}{2}}}{\tV_1\tV_2\tilde{\alpha}},\,\frac{q^{\frac{M_2}{2}}}{
V_1V_2}\right),\\\el
& V = \frac{1}{W} \left(\begin{array}{cccc}
\frac{1}{i\xi}\!\left[{\scriptstyle
U_{z}\xi^{2}-V_{z}+}\frac{\tV_{z}V_{z}-\tU_{z}U_{z}\xi^{2}}{z_{12}}\right] &
V_{z}-U_{z} & i\xi U_{z} & -V_{z}\\
\tU_{z}-\tV_{z} & \frac{i}{\xi}\bigl(\tV_{z}-\tU_{z}\xi^{2}\bigr) & \tV_{z} &
i\tU_{z}\xi\\
\tV_{z}-\tU_{z} & \frac{i}{\xi}\!\left[{\scriptstyle
\tU_{z}\xi^{2}-\tV_{z}+}\frac{\tV_{z}V_{z}-\tU_{z}U_{z}\xi^{2}}{z_{12}}\right] &
-\tV_{z} & -i\tU_{z}\xi\\
\frac{i}{\xi}\left(V_{z}-U_{z}\xi^{2}\right) & V_{z}-U_{z} & iU_{z}\xi & -V_{z}
\end{array}\right),\end{align}
here 
\begin{align}
& W=\tV_{z}V_{z}-\tU_{z}U_{z}\xi^{2}, && U_z = z_{12} - U_1^2 U^2_2, && \tU_z =
z_{12} - \tU_1^2 \tU^2_2,
\end{align}
plus similar expressions for $V_z$. 


\paragraph{Rational limit.}

The matrices $B^+$ \eqref{sIIBP} and $B^-$ \eqref{sIIBM} in the $q\to1+h$
($h\to0$) limit become
\begin{align}
B^+ & = 2 h\,\bar{k}_1 
{\frac{\delta u-\frac{\delta M}{2}-k_2+n+1}{\delta u-\frac{M}{2}+K+1}}
\left(
\begin{array}{cccc}
 0 & 0 & 0 & 0 \\
 0 & 0 & 0 & 0 \\
-k_2c_2/U_1 & 0 & d_2/U_1 & 0 \\
k_2a_2U_1/\alpha\tilde\alpha & 0 & -b_2U_1/\alpha\tilde\alpha & 0 
\end{array}
\right),\\
B^{-} &=2 h\,\bar{k}_2
{\frac{\delta u+\frac{\delta M}{2}+k_2-n+1}{\delta u-\frac{M}{2}+K+1}}
\left(
\begin{array}{cccc}
 0 & 0 & 0 & 0 \\
 0 & 0 & 0 & 0 \\
 0 & -k_1c_1 & 0 & d_1 \\
 0 & k_1a_1/\alpha\tilde\alpha & 0 & -b_1/\alpha\tilde\alpha
\end{array}
\right). 
\end{align}
%

%
The matrices $A$ \eqref{sIIA} and $B$ \eqref{sIIB} in the $q\to1$ limit become
\begin{align}
A &= 
\left(
\begin{array}{cccc}
 {\scriptstyle-(M_2-K+n) g_2 \gamma_2} & {\scriptstyle(M_1-n) g_1 U_2 \gamma_1}
& -\frac{\alpha  g_2 (\xmm-\xpp)}{\gamma_2 \xmm} & \frac{\alpha  g_1 U_2
(\xm-\xp)}{\gamma_1 \xm} \\
 -\frac{i (M_2-K+n) \tilde{\alpha}  g_2 \gamma_2}{\xpp} & \frac{i (M_1-n)
\tilde{\alpha}  g_1 \gamma_1}{U_2 \xp} & -\frac{i \alpha  \tilde{\alpha}  g_2
(\xmm-\xpp)}{\gamma_2} & \frac{i \alpha  \tilde{\alpha}  g_1 (\xm-\xp)}{U_2
\gamma_1} \\
 \frac{i (M_2-K+n) g_2 \gamma_2}{\alpha  \xpp} & -\frac{i (M_1-n) g_1
\gamma_1}{\alpha  U_2 \xp} & \frac{i g_2 (\xmm-\xpp)}{\gamma_2} & -\frac{i g_1
(\xm-\xp)}{U_2 \gamma_1} \\
 -\frac{(M_2-K+n) g_2 \gamma_2}{\alpha  \tilde{\alpha} } & \frac{(M_1-n) g_1 U_2
\gamma_1}{\alpha  \tilde{\alpha} } & -\frac{g_2 (\xmm-\xpp)}{\tilde{\alpha} 
\gamma_2 \xmm} & \frac{g_1 U_2 (\xm-\xp)}{\tilde{\alpha}  \gamma_1 \xm}
\end{array}
\right),\\
B &= 
\left(
\begin{array}{cccc}
 {\scriptstyle -(M_2-k_2) g_2 U_1 \gamma_2} & {\scriptstyle (M_1-k_1) g_1
\gamma_1} & -\frac{\alpha  g_2 U_1 (\xmm-\xpp)}{\gamma_2 \xmm} & \frac{\alpha 
g_1 (\xm-\xp)}{\gamma_1 \xm} \\
 -\frac{i (M_2-k_2) \tilde{\alpha}  g_2 \gamma_2}{U_1 \xpp} & \frac{i (M_1-k_1)
\tilde{\alpha}  g_1 \gamma_1}{\xp} & -\frac{i \alpha  \tilde{\alpha}  g_2
(\xmm-\xpp)}{U_1 \gamma_2} & \frac{i \alpha  \tilde{\alpha}  g_1
(\xm-\xp)}{\gamma_1} \\
 \frac{i (M_2-k_2) g_2 \gamma_2}{\alpha  U_1 \xpp} & -\frac{i (M_1-k_1) g_1
\gamma_1}{\alpha  \xp} & \frac{i g_2 (\xmm-\xpp)}{U_1 \gamma_2} & -\frac{i g_1
(\xm-\xp)}{\gamma_1} \\
 -\frac{(M_2-k_2) g_2 U_1 \gamma_2}{\alpha  \tilde{\alpha} } & \frac{(M_1-k_1)
g_1 \gamma_1}{\alpha  \tilde{\alpha} } & -\frac{g_2 U_1
(\xmm-\xpp)}{\tilde{\alpha}  \gamma_2 \xmm} & \frac{g_1
(\xm-\xp)}{\tilde{\alpha}  \gamma_1 \xm}
\end{array}
\right).
\end{align}
The notation used in here is $g_i = \sqrt{\frac{g}{M_i}}$ and
$U_i=\sqrt{\frac{x^{+}_i}{x^{-}_i}}$. 

It might seem that the matrices $B^+$ and $B^-$ do not contribute in the $q\to1$
limit as they are of order $\mathcal{O}(h)$, however the combinations
$A^{-1}B^{+}$ and $A^{-1}B^{-}$ in \eqref{eqY} are of order $\mathcal{O}(1)$,
thus are defined correctly. We do not spell out the explicit expression of
$A^{-1}$ in the $q\to1$ limit as it is quite sizy and also not much
illuminative.


\subsection{Subspace III}\label{app:A2}

The coefficients' matrices in the expressions \eqref{S3eq2} 
\begin{align}
G^{op}(n,K\!-n)\,\mathscr{Z}^{k_1,k_2}_{n} &=
\mathscr{Y}^{k_1,k_2}_{n}\,G(k_1,k_2),\el
H^{op}(n,K-n)\,\mathscr{Z}^{k_1,k_2}_{n} &=
\overline{\mathscr{Y}}^{k_1,k_2}_{n}\,H(k_1,k_2),\nonumber
\end{align}
are
\begin{align}
G^{op} &= \left(\scriptstyle{
\begin{array}{cccccc}
 \frac{q^{\frac{M_2}{2}\!-\!K\!+n} [M_1\!-n]_q\, a_1}{\tU_2 \tV_2} & 0 &
\frac{q^{\frac{M_2}{2}\!-\!K\!+n} b_1}{\tU_2 \tV_2} & 0 & 0 & -b_2 \\
 \scriptstyle{{[M_2\!-\!K\!+n]_q} a_2} & b_2 & 0 & 0 &
\frac{q^{\frac{M_2}{2}\!-\!K\!+n} b_1}{\tU_2 \tV_2} & 0 \\
 0 & \!\!\!\!\!\!\!\frac{q^{\frac{M_2}{2}\!-\!K\!+n} {[M_1\!-\!n]_q} a_1}{\tU_2
\tV_2} & 0 & \!\!\!\!\frac{q^{\frac{M_2}{2}\!-\!K\!+n} b_1}{\tU_2 \tV_2} & 0 &
\!\!\!\!\!\!\!\!\!\!\!\scriptstyle{[M_2\!-\!K\!+\!n]_q a_2} \\
 0 & 0 & \!\!\!\!\!\!\!\!\scriptstyle{[M_2\!-\!K\!+\!n]_q a_2} & b_2 &
\!\!\!\!\frac{q^{\frac{M_2}{2}\!-\!K\!+n}{[M_1\!-n]_q} a_1}{-\tU_2 \tV_2} & 0
\end{array}}\right),\\\el
G &= \left(
\begin{array}{cccccc}
 \scriptstyle{[M_1-k_1]_q a_1} & 0 & b_1 & 0 & 0 & \frac{q^{\frac{M_1}{2}-k_1}
b_2}{- \tU_1 \tV_1} \\
 \frac{q^{\frac{M_1}{2}-k_1} [M_2-k_2]_q a_2}{ \tU_1 \tV_1} &
\frac{q^{\frac{M_1}{2}-k_1} b_2}{\tU_1 \tV_1} & 0 & 0 & b_1 & 0 \\
 0 & \scriptstyle{[M_1-k_1]_q a_1} & 0 & b_1 & 0 &
\!\!\!\!\!\!\!\!\frac{q^{\frac{M_1}{2}-k_1} [M_2-k_2] a_2}{ \tU_1 \tV_1} \\
 0 & 0 & \!\!\!\!\!\!\!\frac{q^{\frac{M_1}{2}-k_1} [M_2-k_2]_q a_2}{ \tU_1
\tV_1} & \frac{q^{\frac{M_1}{2}-k_1} b_2}{ \tU_1 \tV_1} &
\scriptstyle{-[M_1-k_1]_q a_1} & 0
\end{array}
\right),\label{sIIIG}
\end{align}
and
\begin{align}
H^{op} &= \left(\scriptstyle{
\begin{array}{cccccc}
 \frac{[n]_q c_1}{U_2} & 0 & -\frac{d_1}{U_2} & 0 & -\frac{q^{n-\frac{M_1}{2}}
d_2}{V_1} & 0 \\
 \frac{q^{n-\frac{M_1}{2}} [K-n]_q c_2}{V_1} & \frac{q^{n-\frac{M_1}{2}}
d_2}{-V_1} & 0 & 0 & 0 & \frac{d_1}{U_2} \\
 0 & \frac{[n]_q c_1}{U_2} & 0 & -\frac{d_1}{U_2} &
\!\!-\frac{q^{n-\frac{M_1}{2}} [K-n]_q c_2}{V_1} & 0 \\
 0 & 0 & \frac{q^{n-\frac{M_1}{2}} [K-n]_q c_2}{V_1} & \frac{q^{n-\frac{M_1}{2}}
d_2}{-V_1} & 0 & \frac{[n]_q c_1}{U_2}
\end{array}
}\right),\\\el
H &= \left(\scriptstyle{
\begin{array}{cccccc}
 \frac{q^{k_2-\frac{M_2}{2}} [k_1]_q c_1}{V_2} & 0 & \frac{q^{k_2-\frac{M_2}{2}}
d_1}{-V_2} & 0 & -d_2 \tU_1 & 0 \\
 {[k_2]}_q c_2 \tU_1 & -d_2 \tU_1 & 0 & 0 & 0 & \frac{q^{k_2-\frac{M_2}{2}}
d_1}{V_2} \\
 0 & \frac{q^{k_2-\frac{M_2}{2}} [k_1]_q c_1}{V_2} & 0 &
\frac{q^{k_2-\frac{M_2}{2}} d_1}{-V_2} & -[k_2]_q c_2 \tU_1 & 0 \\
 0 & 0 & [k_2]_q c_2 \tU_1 & -d_2 \tU_1 & 0 & \frac{q^{k_2-\frac{M_2}{2}}
[k_1]_q c_1}{V_2}
\end{array}}\right).
\end{align}
Their affine counterparts $\wtl G$, $\wtl G^{op}$ and $\wtl H$, $\wtl H^{op}$
are obtained by the replacing non-affine (or affine) parameter to affine (or
non-affine) ones.
The matrix $\overline{\mathscr{Y}}^{k_1,k_2}_n$ is a slightly modified version
of ${\mathscr{Y}}^{k_1,k_2}_n$,
\begin{eqnarray}
\overline{\mathscr{Y}}^{k_1,k_2}_n &\equiv& \begin{pmatrix}
  (\mathscr{Y}^{k_1-1,k_2}_{n-1})^1_1 & (\mathscr{Y}^{k_1,k_2-1}_{n-1})^1_2 &
(\mathscr{Y}^{k_1-1,k_2}_{n-1})^1_3 & (\mathscr{Y}^{k_1,k_2-1}_{n-1})^1_4 \\
  (\mathscr{Y}^{k_1-1,k_2}_{n})^2_1   & (\mathscr{Y}^{k_1,k_2-1}_{n})^2_2   &
(\mathscr{Y}^{k_1-1,k_2}_{n})^2_3   & (\mathscr{Y}^{k_1,k_2-1}_{n})^2_4   \\
  (\mathscr{Y}^{k_1-1,k_2}_{n-1})^3_1 & (\mathscr{Y}^{k_1,k_2-1}_{n-1})^3_2 &
(\mathscr{Y}^{k_1-1,k_2}_{n-1})^3_3 & (\mathscr{Y}^{k_1,k_2-1}_{n-1})^3_4 \\
  (\mathscr{Y}^{k_1-1,k_2}_{n})^4_1   & (\mathscr{Y}^{k_1,k_2-1}_{n})^4_2   &
(\mathscr{Y}^{k_1-1,k_2}_{n})^4_3   & (\mathscr{Y}^{k_1,k_2-1}_{n})^4_4
\end{pmatrix}.
\end{eqnarray}
\\

\noindent The coefficient matrices in \eqref{eqZ},
$A\,\mathscr{Z}^{k_1,k_2}_{n} = \check{\mathscr{Y}}^{k,l}_n B,\nonumber$ are
\begin{align}
A &= \left(
\begin{array}{cccccc}
 -\frac{[M_1-n] \mathcal{A}_3}{U_2 V_2} & 0 & \frac{\mathcal{A}_1}{U_2 V_2} & 0
& 0 & q_2 \tilde{z}_2 \\
 0 & -q_2 \tilde{z}_2 & 0 & 0 & \frac{\mathcal{A}_1}{U_2 V_2} & 0 \\
 0 & -\frac{[M_1-n] \mathcal{A}_3}{U_2 V_2} & 0 & \frac{\mathcal{A}_1}{U_2 V_2}
& 0 & 0 \\
 -\frac{[n]_q \mathcal{A}_2}{U_2 V_1} & 0 & -\frac{\mathcal{A}_4}{U_2 V_1} & 0 &
\frac{\tilde{g}^2 q_1}{g^2 \tilde{z}_2} & 0 \\
 0 & \frac{\tilde{g}^2 q_1}{g^2 \tilde{z}_2} & 0 & 0 & 0 &
\frac{\mathcal{A}_4}{U_2 V_1} \\
 0 & -\frac{[n]_q \mathcal{A}_2}{U_2 V_1} & 0 & -\frac{\mathcal{A}_4}{U_2 V_1} &
0 & 0
\end{array}
\right),
\end{align}
\begin{align}
&A^{-1} = \scriptstyle{\left(
\begin{array}{cccccc}
 -\frac{U_2 V_2}{\mathcal{A}_0 \mathcal{A}_4^{-1}} & \frac{\tilde{g}^2 q_1 U_2^2
V_1 V_2}{g^2 \mathcal{A}_0 \tilde{z}_2} & 0 & -\frac{U_2 V_1}{\mathcal{A}_0
\mathcal{A}_1^{-1}} & \frac{q_2 U_2^2 V_1 V_2 \tilde{z}_2}{\mathcal{A}_0} & 0 \\
 0 & 0 & -\frac{U_2 V_2}{\mathcal{A}_0 \mathcal{A}_4^{-1}} & 0 & 0 & -\frac{U_2
V_1}{\mathcal{A}_0 \mathcal{A}_1^{-1}}\\
 \frac{[n]_q U_2 V_2}{\mathcal{A}_0 \mathcal{A}_2^{-1}} & \frac{\tilde{g}^2
[M_1-n] q_1 U_2^2 V_1 V_2}{g^2 \mathcal{A}_0 \mathcal{A}_1 \mathcal{A}_3^{-1}
\tilde{z}_2} & \frac{\tilde{g}^2 q_1 q_2 U_2^3 V_1 V_2^2}{-g^2 \mathcal{A}_0
\mathcal{A}_1} & \frac{[M_1-n] U_2 V_1}{-\mathcal{A}_0 \mathcal{A}_3^{-1}} &
\frac{[n]_q q_2 U_2^2 V_1 V_2 \tilde{z}_2}{-\mathcal{A}_0 \mathcal{A}_2^{-1}
\mathcal{A}_4} & \frac{\tilde{g}^2 q_1 q_2 U_2^3 V_1^2 V_2}{-g^2 \mathcal{A}_0
\mathcal{A}_4} \\
 0 & 0 & \frac{[n]_q U_2 V_2}{\mathcal{A}_0 \mathcal{A}_2^{-1}} & 0 & 0 &
-\frac{[M_1-n] U_2 V_1}{\mathcal{A}_0 \mathcal{A}_3^{-1}} \\
 0 & \frac{U_2 V_2}{\mathcal{A}_1} & -\frac{q_2 U_2^2 V_2^2
\tilde{z}_2}{\mathcal{A}_0 \mathcal{A}_1 \mathcal{A}_4^{-1}} & 0 & 0 &
-\frac{q_2 U_2^2 V_1 V_2 \tilde{z}_2}{\mathcal{A}_0} \\
 0 & 0 & \frac{\tilde{g}^2 q_1 U_2^2 V_1 V_2}{g^2 \mathcal{A}_0 \tilde{z}_2} & 0
& \frac{U_2 V_1}{\mathcal{A}_4} & \frac{\tilde{g}^2 q_1 U_2^2 V_1^2}{g^2
\mathcal{A}_0 \mathcal{A}_1^{-1} \mathcal{A}_4 \tilde{z}_2}
\end{array}
\right)},
\end{align}
here we have defined $\tl z_i = \frac{\tl g \alpha \tl \alpha}{g} z_i$ and 
$\mathcal{A}_{0} = [n]_q \,\mathcal{A}_{1}\mathcal{A}_{2} +
[M_1-n]_q\,\mathcal{A}_{3}\mathcal{A}_{4}$ where
\begin{align}
  \mathcal{A}_{1} &= b_1\tilde{a}_2 U_2^2 V_2^2 - a_2 \tilde{b}_1,
 &\mathcal{A}_{2} &= c_2\tilde{c}_1 U_2^2 - c_1 V_1^2 \tilde{c}_2,\el
  \mathcal{A}_{3} &= a_2 \tilde{a}_1 - a_1\tilde{a}_2 U_2^2 V_2^2,
 &\mathcal{A}_{4} &= d_1\tilde{c}_2 V_1^2 - c_2\tilde{d}_1
U_2^2.\label{s3Acoefs}
\end{align}
\begin{align}
B &= \left(\scriptstyle{
\begin{array}{cccccc}
 \scriptstyle{-[M_1-k_1]_q q_2 \mathcal{B}_3} & 0 & q_2 \mathcal{B}_2 & 0 & 0 &
-\frac{q_3 \mathcal{B}_1}{q_1 U_1 V_1} \\
 -\frac{[M_2-k_2]_q q_3 \mathcal{B}_7}{q_1 U_1 V_1} & \frac{q_3
\mathcal{B}_1}{q_1 U_1 V_1} & 0 & 0 & q_2 \mathcal{B}_2 & 0 \\
 0 & \!\!\!\!\scriptstyle{-[M_1-k_1]_q q_2 \mathcal{B}_3} & 0 & q_2
\mathcal{B}_2 & 0 & \!\!\!\!\!\!-\frac{[M_2-k_2]_q q_3 \mathcal{B}_7}{q_1 U_1
V_1} \\
 0 & 0 & \!\!\!\!-\frac{[M_2-k_2]_q q_3 \mathcal{B}_7}{q_1 U_1 V_1} & \frac{q_3
\mathcal{B}_1}{q_1 U_1 V_1} & \scriptstyle{[M_1-k_1]_q q_2 \mathcal{B}_3} & 0 \\
 -\frac{[k_1]_q q_3 \mathcal{B}_4}{V_1 V_2} & 0 & -\frac{q_3 \mathcal{B}_5}{V_1
V_2} & 0 & -\frac{\mathcal{B}_6}{U_1 V_1} & 0 \\
 -\frac{[k_2]_q \mathcal{B}_8}{U_1 V_1} & -\frac{\mathcal{B}_6}{U_1 V_1} & 0 & 0
& 0 & \frac{q_3 \mathcal{B}_5}{V_1 V_2} \\
 0 & -\frac{[k_1]_q q_3 \mathcal{B}_4}{V_1 V_2} & 0 & -\frac{q_3
\mathcal{B}_5}{V_1 V_2} & \frac{[k_2]_q \mathcal{B}_8}{U_1 V_1} & 0 \\
 0 & 0 & -\frac{[k_2]_q \mathcal{B}_8}{U_1 V_1} & -\frac{\mathcal{B}_6}{U_1 V_1}
& 0 & -\frac{[k_1]_q q_3 \mathcal{B}_4}{V_1 V_2}
\end{array}}
\right),
\end{align}
and we are using the shorthand notation $q_1=q^{n-\frac{M_1}{2}}$,
$q_2=q^{K-n-\frac{M_2}{2}}$, $q_3=q^{k_2-\frac{M_2}{2}}$ and
\begin{align}
\mathcal{B}_{1} &= b_2 \tilde{a}_2 U_1^2 V_1^2 - a_2 \tilde{b}_2,
&\mathcal{B}_{2} &= b_1 \tilde{a}_2 - a_2 \tilde{b}_1,\el
\mathcal{B}_{3} &= a_2 \tilde{a}_1 - a_1 \tilde{a}_2, &\mathcal{B}_{4} &= c_2
\tilde{c}_1 V_2^2 - c_1 \tilde{c}_2 V_1^2,\el
\mathcal{B}_{5} &= d_1 \tilde{c}_2 V_1^2 - c_2 \tilde{d}_1 V_2^2,
&\mathcal{B}_{6} &= d_2 \tilde{c}_2 V_1^2 - c_2 \tilde{d}_2 U_1^2,\el
\mathcal{B}_{7} &= a_2 \tl a_2 (1-U_1^2 V_1^2), &\mathcal{B}_{8} &= c_2 \tl c_2
(U_1^2-V_1^2).\label{s3Bcoefs}
\end{align}

\noindent The matrix $\check{\mathscr{Y}}^{k_1,k_2}_n$ is defined as
\begin{align}
\check{\mathscr{Y}}^{k_1,k_2}_n = 
\left(
\begin{array}{cc}
 {{\mathscr{Y}}^{k_1,k_2}_n} & 0 \\
 0 & {\overline{\mathscr{Y}}^{k_1,k_2}_n} 
\end{array}
\right),
\end{align}
where only first three rows of both ${\mathscr{Y}}^{k_1,k_2}_n$ and
$\overline{\mathscr{Y}}^{k_1,k_2}_n$ are taken.


\paragraph{Rational limit.}

In the rational limit $q\to1$ the coefficients \eqref{s3Acoefs} and
\eqref{s3Bcoefs} acquire quite compact expressions
\begin{align}
\frac{\mathcal{A}_1}{\alpha\,\tl\alpha} = \alpha\,\tl\alpha \, \mathcal{A}_4 &=
i \sqrt{\frac{g}{M_1}} \sqrt{\frac{g}{M_2}}
\frac{\left(\xm-\xp\right)\left(1-\xm \xmm\right) \gamma
_2}{\xm \xmm \gamma_1},\el
\frac{\mathcal{A}_3}{\tl\alpha} = \tl\alpha \, \mathcal{A}_2 &= i
\sqrt{\frac{g}{M_1}} \sqrt{\frac{g}{M_2}} \frac{\left(\xmm-\xp\right) \gamma_1
\gamma_2}{\xmm \xp},
\end{align}
giving
\begin{align}
\mathcal{A}_0 = -\frac{g^2}{\alpha M_2} \frac{\left(1 - \xm \xmm\right)
\left(\xm - \xp\right) \left(\xmm - \xp \right) \gamma_2^2}{\xm (\xmm)^2 \xp},
\end{align}
and also
\begin{align}
\frac{\mathcal{B}_1}{\alpha\,\tl\alpha} = \alpha\,\tl\alpha \, \mathcal{B}_6 &=
i \frac{g}{M_2}\frac{\left(\xmm-\xpp\right) \left(\xp-\xm \xmm \xpp\right)}{\xm
\xmm \xpp},\el
\frac{\mathcal{B}_2}{\alpha\,\tl\alpha} = \alpha\,\tl\alpha \, \mathcal{B}_5 &=
i \sqrt{\frac{g}{M_1}} \sqrt{\frac{g}{M_2}} \frac{\left(\xm-\xp\right)
\left(1-\xm \xpp\right) \gamma_2}{\xm \xpp \gamma_1},\el
\frac{\mathcal{B}_3}{\tl\alpha} = \alpha^2 \tl\alpha \, \mathcal{B}_4 &= -i
\sqrt{\frac{g}{M_1}} \sqrt{\frac{g}{M_2}} \frac{\left(\xp-\xpp\right) \gamma_1
\gamma_2}{\xp \xpp},\el
\frac{\mathcal{B}_7}{\tl\alpha} = \alpha^2 \tl\alpha \, \mathcal{B}_8 &= i
\frac{g}{M_2} \frac{\left(\xm-\xp\right) \gamma_2^2}{\xm \xpp}.
\end{align}
%


\section{Elements of the special cases of the S-matrix}


\subsection{Elements of the fundamental S-matrix}\label{app:fundS}

The fundamental S-matrix for the space III acquires the following form,  
\begin{align}
\S\,\ket{\phi^1\phi^2}&=(\mathscr{Z}^{1,0}_{1})^1_1\ket{\phi^1\phi^2}+(\mathscr{
Z}^{1,0}_{0})^1_1\ket{\phi^2\phi^1}
+(\mathscr{Z}^{1,0}_{1})^5_1\ket{\psi^1\psi^2}+(\mathscr{Z}^{1,0}_{0})^6_1\ket{
\psi^2\psi^1},
\el
\S\,\ket{\phi^2\phi^1}&=(\mathscr{Z}^{0,1}_{1})^1_1\ket{\phi^1\phi^2}+(\mathscr{
Z}^{0,1}_{0})^1_1\ket{\phi^2\phi^1}
+(\mathscr{Z}^{0,1}_{1})^5_1\ket{\psi^1\psi^2}+(\mathscr{Z}^{0,1}_{0})^6_1\ket{
\psi^2\psi^1},
\el
\S\,\ket{\psi^1\psi^2}&=(\mathscr{Z}^{1,0}_{1})^1_5\ket{\phi^1\phi^2}+(\mathscr{
Z}^{1,0}_{0})^1_5\ket{\phi^2\phi^1}
+(\mathscr{Z}^{1,0}_{1})^5_5\ket{\psi^1\psi^2}+(\mathscr{Z}^{1,0}_{0})^6_5\ket{
\psi^2\psi^1},
\el
\S\,\ket{\psi^2\psi^1}&=(\mathscr{Z}^{0,1}_{1})^1_6\ket{\phi^1\phi^2}+(\mathscr{
Z}^{0,1}_{0})^1_6\ket{\phi^2\phi^1}
+(\mathscr{Z}^{0,1}_{1})^5_6\ket{\psi^1\psi^2}+(\mathscr{Z}^{0,1}_{0})^6_6\ket{
\psi^2\psi^1}.
\end{align}
In order to find these coefficients $\mathscr{Z}$ it is sufficient to consider
the first relation of \eqref{S3eq2} and its affine counterpart only. 
In fact, the constraints read as follows,
\begin{align}
\begin{pmatrix}(G^{op})^2_1&(G^{op})^2_5\\(\wtl G^{op})^2_1&(\wtl
G^{op})^2_5\end{pmatrix}\!\!(1,0)
\begin{pmatrix}(\mathscr{Z}^{1,0}_{1})^1_1&(\mathscr{Z}^{1,0}_{1})^1_5\\
(\mathscr{Z}^{1,0}_{1})^5_1&(\mathscr{Z}^{1,0}_{1})^5_5\end{pmatrix}
&=
(\mathscr{Y}^{1,0}_{1})^2_2
\begin{pmatrix}(G)^2_1&(G)^2_5\\(\wtl G)^2_1&(\wtl
G)^2_5\end{pmatrix}\!\!(1,0)\,,
\el
\begin{pmatrix}(G^{op})^2_1&(G^{op})^2_5\\(\wtl G^{op})^2_1&(\wtl
G^{op})^2_5\end{pmatrix}\!\!(1,0)
\begin{pmatrix}(\mathscr{Z}^{0,1}_{1})^1_1&(\mathscr{Z}^{0,1}_{1})^1_6\\
(\mathscr{Z}^{0,1}_{1})^5_1&(\mathscr{Z}^{0,1}_{1})^5_6\end{pmatrix}
&=
(\mathscr{Y}^{0,1}_{1})^2_1
\begin{pmatrix}(G)^1_1&(G)^1_6\\(\wtl G)^1_1&(\wtl
G)^1_6\end{pmatrix}\!\!(0,1)\,,
\el
\begin{pmatrix}(G^{op})^1_1&(G^{op})^1_6\\(\wtl G^{op})^1_1&(\wtl
G^{op})^1_6\end{pmatrix}\!\!(0,1)
\begin{pmatrix}(\mathscr{Z}^{1,0}_{0})^1_1&(\mathscr{Z}^{1,0}_{0})^1_5\\
(\mathscr{Z}^{1,0}_{0})^6_1&(\mathscr{Z}^{1,0}_{0})^6_5\end{pmatrix}
&=
(\mathscr{Y}^{1,0}_{0})^1_2
\begin{pmatrix}(G)^2_1&(G)^2_5\\(\wtl G)^2_1&(\wtl
G)^2_5\end{pmatrix}\!\!(1,0)\,,
\el
\begin{pmatrix}(G^{op})^1_1&(G^{op})^1_6\\(\wtl G^{op})^1_1&(\wtl
G^{op})^1_6\end{pmatrix}\!\!(0,1)
\begin{pmatrix}(\mathscr{Z}^{0,1}_{0})^1_1&(\mathscr{Z}^{0,1}_{0})^1_6\\
(\mathscr{Z}^{0,1}_{0})^6_1&(\mathscr{Z}^{0,1}_{0})^6_6\end{pmatrix}
&=itself
(\mathscr{Y}^{0,1}_{0})^1_1
\begin{pmatrix}(G)^1_1&(G)^1_6\\(\wtl G)^1_1&(\wtl G)^1_6\end{pmatrix}\!\!(0,1)
\,.
\end{align}
It is easy to solve these relations for $\mathscr{Z}$ and we find 
them to agree with \cite{BK}. 
For the completeness, we have listed the relations of our elements $\mathscr{Z}$
to those of \cite{BK}%
\footnote{We remind that our $x^\pm$ parameterization is based on the one of 
\cite{BGM} which are related to those of \cite{BK} by $x^\pm_{\text{[35]}}=g\tl g^{-1}
(x^\pm_{\text{[10]}}+\xi)$. This point must be taken into account when
performing the concrete comparison.}
\begin{align}
\begin{pmatrix}(\mathscr{Z}^{1,0}_{1})^1_1&(\mathscr{Z}^{1,0}_{1})^1_5\\
(\mathscr{Z}^{1,0}_{1})^5_1&(\mathscr{Z}^{1,0}_{1})^5_5\end{pmatrix}
=
\begin{pmatrix}(\mathscr{Z}^{0,1}_{0})^1_1&(\mathscr{Z}^{0,1}_{0})^1_6\\
(\mathscr{Z}^{0,1}_{0})^6_1&(\mathscr{Z}^{0,1}_{0})^6_6\end{pmatrix}
&=
\frac{1}{A_{12}}
\begin{pmatrix}
\frac{A_{12}-B_{12}}{q+q^{-1}}&-\frac{F_{12}}{q+q^{-1}}\\
\frac{C_{12}}{q+q^{-1}}&-\frac{D_{12}-E_{12}}{q+q^{-1}}
\end{pmatrix},
\el
\begin{pmatrix}(\mathscr{Z}^{0,1}_{1})^1_1&(\mathscr{Z}^{0,1}_{1})^1_6\\
(\mathscr{Z}^{0,1}_{1})^5_1&(\mathscr{Z}^{0,1}_{1})^5_6\end{pmatrix}
&=
\frac{1}{A_{12}}
\begin{pmatrix}
\frac{q^{-1} A_{12}+qB_{12}}{q+q^{-1}}&\frac{q F_{12}}{q+q^{-1}}\\ -\frac{q
C_{12}}{q+q^{-1}}&-\frac{q^{-1}D_{12}+qE_{12}}{q+q^{-1}}
\end{pmatrix},
\el
\begin{pmatrix}(\mathscr{Z}^{0,1}_{0})^1_1&(\mathscr{Z}^{0,1}_{0})^1_6\\
(\mathscr{Z}^{0,1}_{0})^6_1&(\mathscr{Z}^{0,1}_{0})^6_6\end{pmatrix}
&=
\frac{1}{A_{12}}
\begin{pmatrix}
\frac{qA_{12}+q^{-1}B_{12}}{q+q^{-1}}&\frac{q^{-1} F_{12}}{q+q^{-1}}\\
-\frac{q^{-1} C_{12}}{q+q^{-1}}&-\frac{qD_{12}+q^{-1}E_{12}}{q+q^{-1}}
\end{pmatrix}.
\end{align}
%


\subsection{Elements of the S-matrix S\texorpdfstring{$_{Q1}$}{SQ1}}\label{app:SQ1}

Here we list the explicit forms of the coefficients of the matrix $S_{Q1}$.


\paragraph{Subspace II.} 

First we give the coefficients of the matrix
$\mathscr{Y}$ in the case of a bound state scattering with a fundamental
particle. There are four different combinations of the parameters $k_1,k_2,n$
that contribute. Thus we have to consider the case where $k_2=0$ and $k_1=n=k$
leading to
\begin{align}
&(\mathscr{Y}^{k,0}_{k-1})^1_1 = q^{\frac{1}{2}+k} U_2 V_2
\frac{x^-_1-x^-_2}{x^-_1-x^+_2} \frac{z_{12}-q^{Q-2k-1}}{z_{12}-q^{Q-1}}, &&\!
(\mathscr{Y}^{k,0}_{k-1})^2_2 =  \frac{1}{q^{\frac{Q}{2}}U_1V_1}\frac{x^+_1 -
x^+_2}{x^-_1 - x^+_2},\nonumber\\
&(\mathscr{Y}^{k,0}_{k-1})^1_2 =  q^{\frac{1 -Q}{2}} \frac{[Q -
k]_q}{\sqrt{[Q]_q}} \frac{x^-_2-x^+_2}{x^-_1-x^+_2}
\frac{U_2 V_2}{U_1 V_1}\frac{\gamma_1}{\gamma_2}, &&\!
(\mathscr{Y}^{k,0}_{k-1})^2_1 = \frac{1}{\sqrt{[Q]_q}} \frac{x^-_1 -
x^+_1}{x^-_1 - x^+_2}\frac{\gamma_2}{\gamma_1},\nonumber\\
&(\mathscr{Y}^{k,0}_{k-1})^1_4 =\frac{q^{\frac{1-Q}{2}}\alpha}{\sqrt{[Q]_q}}
\frac{U_2 V_2}{U_1V_1} 
\frac{\![x^-_1-x^+_1]\![x^-_2-x^+_2]\![x^-_2-x^+_1]}{
(x^-_1-x^+_2)(x^-_1x^-_2-1)\gamma_1\gamma_2}, && \!
(\mathscr{Y}^{k,0}_{k-1})^4_2=(\mathscr{Y}^{k,0}_{k-1})^2_4=0,\\
&
(\mathscr{Y}^{k,0}_{k-1})^4_1 = \frac{q^{-Q}[k]_q}{\sqrt{[Q]_q}}
\frac{x^+_1-x^-_2}{(x^-_1-x^+_2)(1-x^-_1x^-_2)}\frac{x^-_1}{x^+_1}\frac{
\gamma_1\gamma_2}{\alpha}, &&\!
(\mathscr{Y}^{k,0}_{k-1})^4_4 =
\frac{q^{-\frac{Q}{2}}}{U_1V_1}\frac{x^+_1-x^-_1}{x^-_1-x^+_2}\frac{1-x^-_1x^+_2
}{1-x^-_1x^-_2}.\nonumber
\end{align}
Next we have three elements corresponding to $k_2=1$ and $k_1+1=n=k$ giving
\begin{align}
&(\mathscr{Y}^{k,0}_{k-1})^1_1 = q^{\frac{1}{2} - Q} U_2V_2
\frac{x^-_1-x^-_2}{x^-_1-x^+_2} \frac{(q^{2(k+1)}
-q^{2Q})z_{12}}{z_{12}-q^{Q-1}},&&
(\mathscr{Y}^{k,0}_{k-1})^2_1 =
\frac{q^{1+k-Q}}{\sqrt{[Q]_q}}\frac{x^-_1-x^+_1}{x^-_1-x^+_2}\frac{\gamma_2}{
\gamma_1},\el
& (\mathscr{Y}^{k,0}_{k-1})^4_1 = \frac{[Q - k - 1]_q}{q^{Q-k-1}\sqrt{[Q]_q}}
\frac{x^-_2-x^+_1}{(x^-_1-x^+_2)(1-x^-_1x^-_2)}\frac{x^-_1}{x^+_1}\frac{
\gamma_1\gamma_2}{\alpha}.
\end{align}
Then we have another three scattering entries for $k_2=0$ and $k_1=n+1=k$
contributing
\begin{align}
&(\mathscr{Y}^{k,0}_{k-1})^1_1 = q^{\frac{1}{2} + Q} U_2 V_2 
\frac{x^-_1-x^-_2}{x^-_1-x^+_2} \frac{1 - q^{-2 k}}{q^Q - q z_{12}},&&
(\mathscr{Y}^{k,0}_{k-1})^1_2 =
q^{\frac{1+Q-2k}{2}}\frac{[k]_q}{\sqrt{[Q]_q}}\frac{x^-_2-x^+_2}{x^-_1-x^+_2}
\frac{U_2 V_2}{U_1 V_1}\frac{\gamma_1}{\gamma_2}, \nonumber \\
& (\mathscr{Y}^{k,0}_{k-1})^1_4 = -q^{-k}(\mathscr{Y}^{k,0}_{k})^1_4.
\end{align}
Finally, there is one element with $k_2=1$ and $k_1=n=k-1$ providing the last
element
\begin{align}
(\mathscr{Y}^{k-1,1}_{k-1})^1_1 = q^{\frac{1}{2}-k}U_2V_2
\frac{x^-_1-x^-_2}{x^-_1-x^+_2}\frac{q^{2k} - q^{1+Q} z_{12}}{q^Q - q z_{12}}.
\end{align}
%


\paragraph{Subspace III.}

There are 36 elements of the matrix $\mathscr{Z}$ that need be determined. 
As mentioned in Section \ref{sec:SpecialCases}, it follows that
(\ref{eqn;SQ1XtoZ}) becomes
\begin{align}
\S\,\stateC{k-1,1}_6 = \mathscr{D}\left(\stateC{k,0}_5 +
q\,\stateC{k-1,1}_6\right) - q\,\S\,\stateC{k,0}_5.
\end{align}
Acting with the S-matrix on both sides of the equations (\ref{eqn;SQ1ZtoZ}) and
using its invariance property allows us to express the elements of the S-matrix
of the left hand side to the ones on the right hand side. Explicitly we find 
\begin{align}
(\mathscr{Z}^{k-1,1}_k)^{1}_1=&(\mathscr{Z}^{k,0}_k)^{1}_1 [Q - k + 1]_q (q^{2 k
- Q - 2} - q) + (\mathscr{Z}^{k,0}_{k-1})^{1}_1\frac{[Q - k + 1]_q }{[k]_q}
\,,\el
(\mathscr{Z}^{k-1,1}_{k-1})^{1}_1=&(\mathscr{Z}^{k,0}_k)^{1}_1 + 
\frac{[k-1]_q[Q-k+2]_q (q^{2 k-Q-4}-q)-[Q-2
k+1]_q}{[k]_q}(\mathscr{Z}^{k,0}_{k-1})^{1}_1 \,,\el
(\mathscr{Z}^{k-1,1}_{k})^{3}_1=&
(\mathscr{Z}^{k,0}_{k})^{3}_1\frac{[k-1]_q [Q-k]_q q^{2 k-Q-2}-q
[k]_q[Q-k+1]_q}{[k]_q} + 
(\mathscr{Z}^{k,0}_{k-1})^{3}_1\frac{[Q-k]_q}{[k]_q} \,,\el
(\mathscr{Z}^{k-1,1}_{k-1})^{3}_1=&
(\mathscr{Z}^{k,0}_{k-1})^{3}_1 \frac{ [k-2]_q [Q-k+1]_q q^{2 k-Q-4} + q
[k-1]_q[k-Q-2]_q + [2 k-Q-1]_q}{[k]_q} \el
&+(\mathscr{Z}^{k,0}_{k})^{3}_1\frac{[k-1]_q)}{[k]_q} \,,\el
(\mathscr{Z}^{k-1,1}_k)^{5}_1 =& (\mathscr{Z}^{k,0}_k)^{5}_1 \frac{ ([k-1]_q
q^{2k-3-Q}- q[k]_q) [Q-k+1]_q}{[k]_q} \,,\el
(\mathscr{Z}^{k-1,1}_{k-1})^{6}_1 =& (\mathscr{Z}^{k,0}_{k-1})^{6}_1 \frac{
([k-1]_q q^{2k-3-Q}- q[k]_q) [Q-k+1]_q}{[k]_q}\,,\el
(\mathscr{Z}^{k-1,1}_k)^{1}_3=&(\mathscr{Z}^{k,0}_k)^{1}_3\left[\frac{[k]_q[
Q-k+1]_q q^{2 k-Q-2}}{[k-1]_q} - q[Q-k]_q\right] +
(\mathscr{Z}^{k,0}_{k-1})^{1}_3\frac{[Q-k+1]_q}{[k-1]_q} \,,\el
(\mathscr{Z}^{k-1,1}_{k-1})^{1}_3=&(\mathscr{Z}^{k,0}_{k-1})^{1}_3
\left[\frac{[Q-k+2]_q}{q^{Q+4-2k}}+ [k-Q+1]_q - \frac{[k-2]_q
q^{Q-k+1}}{[k-1]_q}\right]+(\mathscr{Z}^{k,0}_k)^{1}_3
\frac{[k]_q}{[k-1]_q} \,, \el
(\mathscr{Z}^{k-1,1}_{k})^{3}_3=&(\mathscr{Z}^{k,0}_{k})^{3}_3 [Q-k]_q(q^{2
k-Q-2}-q) + (\mathscr{Z}^{k,0}_{k-1})^{3}_3
\frac{[Q-k]_q}{[k-1]_q} \,,\nonumber
\end{align}
\begin{align}
(\mathscr{Z}^{k-1,1}_{k-1})^{3}_3=&(\mathscr{Z}^{k,0}_{k-1})^{3}_3
\!\left[\!\frac{[k-2]_q [Q-k+1]_q}{q^{Q-2 k+4}[k-1]_q} + \frac{q[k]_q
[k-Q+1]_q\! - q^2 [2k-Q-1]_q}{[k-1]_q}\!\right]\! +
(\mathscr{Z}^{k,0}_{k})^{3}_3 \,,\el
(\mathscr{Z}^{k-1,1}_k)^{5}_3 =& (\mathscr{Z}^{k,0}_k)^{5}_1 ([Q-k+1]_q q^{2
k-Q-3}-q [Q-k]) \,,\el
(\mathscr{Z}^{k-1,1}_{k-1})^{6}_3 =& (\mathscr{Z}^{k,0}_{k-1})^{6}_1 ([Q-k+1]_q
q^{2 k-Q-3}-q [Q-k]) \,.
\end{align}
Finally, the remaining elements are
\begin{align}
(\mathscr{Z}^{k,0}_k)^{1}_5=& \frac{\alpha}{U_1V_1} \frac{(\xm-\xp) (\xmm-\xpp)
\left[\frac{(\xi\xp+1) [Q-k]_q
(q (\xi
+\xmm)-\xpp-\xi)}{(\xi^2-1)q^{Q}}-[k]_q(\xp-\xpp)\right]}{\gamma_1\gamma_2[k]
_q\sqrt{[Q]} (1-\xm\xmm)(\xm-\xpp)q^{\frac{Q}{2}} }\,, \qquad\qquad\el
(\mathscr{Z}^{k,0}_{k-1})^{1}_5=&\frac{\alpha}{U_1V_1}\frac{(\xm-\xp)(\xmm-\xpp)
[q (\xi +\xmm) 
(\xi \xp+1)-(\xi + \xp)(\xi\xpp+1)]}{\gamma_1\gamma_2(\xi^2-1)
\sqrt{[Q]}(1-\xm\xmm)(\xm-\xpp)q^{k+\frac{Q}{2}} } \,,\el
(\mathscr{Z}^{k,0}_k)^{3}_5=& \frac{\gamma_1}{\gamma_2
q^{\frac{Q}{2}}}\frac{[Q-k]_q}{\sqrt{[Q]_q}} \frac{(\xmm-\xpp) 
[q (\xi + \xm) (\xi +\xmm)-(\xi \xm+1)(\xi\xpp+1)]}{(\xi^2-1) (1-\xm\xmm)  
U_1V_1(\xm-\xpp)} \,,\el
(\mathscr{Z}^{k,0}_{k-1})^{3}_5=&\frac{\gamma_1}{\gamma_2}\frac{[k-1]_q}{\sqrt{[
Q]_q}}\frac{(\xmm-\xpp) 
   [q (\xi + \xm) (\xi +\xmm)-(\xi  \xm+1) (\xi \xpp+1)]}{(\xi
^2-1)(1-\xm\xmm)U_1V_1(\xm-\xpp)q^{k-\frac{Q}{2}}} \,,\el
(\mathscr{Z}^{k,0}_k)^{5}_5=& \frac{ (\xp-\xpp) [(\xi \xm + 1) (\xi
\xpp+1)-q(\xi +\xm)(\xi+\xmm)]}{(\xi ^2-1)(\xm
\xmm-1)(\xm-\xpp)U_1V_1U_2V_2q^{\frac{Q+1}{2}}} \,,\el
(\mathscr{Z}^{k,0}_{k-1})^{6}_5=& \frac{z_{12}(\xmm-\xpp) (\xp(\xi  \xmm +1)(\xi
\xpp + 1)-V_1^4 \xm (\xi + \xmm)
(\xi +\xpp))}{(\xi
^2-1)V_1^2\xpp(\xm\xmm-1)(\xm-\xpp)q^{\frac{1+Q}{2}}}\frac{U_2V_2}{U_1V_1}\,,\el
(\mathscr{Z}^{k,0}_k)^{1}_1=&
\frac{\xmm(\xm-\xp)[Q-k]_Q[(\xi\xm+1)(\xi\xp+1)-V_2^2(\xi+\xm)(\xi+\xp)]}
{(\xi^2-1)\xp z_{12}[Q]_q(\xm \xmm-1)(\xm-\xpp)q^Q}+\nonumber\\
&+\frac{\xm (\xmm \xp-1) (\xp-\xpp)q^{k-2Q}}{\xp (\xm \xmm-1) (\xm-\xpp)}\,,\el
(\mathscr{Z}^{k,0}_{k-1})^{1}_1=&\frac{\xmm [k]_q q^{-k} (\xm-\xp)[(\xi 
\xm+1)(\xi  \xp+1)-V_2^2(\xi +\xm)(\xi+\xp)]}
{(1-\xi ^2)\xp z_{12}[Q]_q(1-\xm\xmm)(\xm-\xpp)}\,,\el
(\mathscr{Z}^{k,0}_k)^{3}_1=& \frac{\gamma_1^2 \xm [k]_q q^{-Q-1}
(\xi\xpp+1)[Q-k]_q [q\xpp(\xi+\xmm)-\xmm(\xi+\xpp)]}
{\alpha(\xi^2-1) \xp \xpp [Q]_q(1-\xm \xmm)(\xm-\xpp)}\,,\el
(\mathscr{Z}^{k,0}_{k-1})^{3}_1=&
\frac{\gamma_1^2 \xm [k-1]_q[k] q^{-k-1} (\xi\xpp+1) [q \xpp (\xi +\xmm)-\xmm
(\xi +\xpp)]}
{\alpha(\xi^2-1)\xp\xpp[Q]_q(1-\xm \xmm)(\xm-\xpp)}\,,\el
(\mathscr{Z}^{k,0}_k)^{5}_1=& \frac{\gamma_1\gamma_2(V_2^2-1)\xm\xmm [k]_q
q^{-Q-\frac{3}{2}}(\xp-\xpp)(\xi +\xpp)(\xi 
   \xpp+1)}
{\alpha (\xi ^2-1) \xp \xpp \sqrt{[Q]_q}(1-\xm \xmm) (\xm-\xpp)
(\xmm-\xpp)U_2V_2}\,,\el
(\mathscr{Z}^{k,0}_{k-1})^{6}_1=& \frac{\gamma_1\gamma_2 \xm [k]_q
q^{-Q-\frac{3}{2}} (\xp-\xpp)}
{\alpha  \xp \sqrt{[Q]_q}(\xm \xmm-1)(\xm-\xpp) U_2V_2}\,,\el
(\mathscr{Z}^{k,0}_k)^{1}_3=& \frac{\alpha  q (1-V_2^2) (\xm-\xp)^2 (\xi +\xmm)
(\xi \xmm+1)}{\gamma_1^2 (\xi ^2-1)[Q]_q
(\xm\xmm-1) (\xm-\xpp)}\,,\el
(\mathscr{Z}^{k,0}_{k-1})^{1}_3=&\frac{\alpha (1-V_2^2) q^{k-3}(\xm-\xp)^2 (\xi
+\xmm)(\xi  \xmm+1)}{
\gamma_1^2(\xi ^2-1)[Q]_q(1-\xm\xmm) (\xm-\xpp)}\,,\nonumber
\end{align}
%
%
\begin{align}
(\mathscr{Z}^{k,0}_k)^{3}_3=&
\frac{q^k (\xm-\xmm) (\xm \xpp-1)}{(\xm \xmm-1)(\xm-\xpp)}+\el
&-
\frac{\xmm [k]_q (\xm-\xp)[V_2^2 (\xi\xm+1)(\xi\xp+1)-(\xi +\xm)(\xi+\xp)]}{(\xi
^2-1)\xp z_{12}[Q]_q(\xm\xmm-1)(\xm-\xpp)}\,,\qquad\qquad\qquad\el
(\mathscr{Z}^{k,0}_{k-1})^{3}_3=&\frac{\xmm [k-1]_q q^{-k}
(\xm-\xp)[V_2^2(\xi\xm+1)(\xi\xp+1)-(\xi+\xm)(\xi+\xp)]}
{(\xi^2-1)\xp z_{12}[Q]_q(\xm \xmm-1)(\xm-\xpp)}\,,\el
(\mathscr{Z}^{k,0}_k)^{5}_3=&\sqrt{\frac{q}{[Q]_q}}\frac{
(V_2^2-1)(\xm-\xp)(1-\xm \xpp)(\xi +\xmm)(\xi\xmm+1)}
{(\xi^2-1)(\xm\xmm-1)(\xm-\xpp)(\xmm-\xpp)}\,,\el
(\mathscr{Z}^{k,0}_{k-1})^{6}_3=& \frac{\gamma_2 q^{-Q-\frac{1}{2}} (\xm-\xp)
(\xm\xpp-1)}
{\gamma_1 \sqrt{[Q]_q}(\xm \xmm-1)(\xm-\xpp)U_2V_2}.
\end{align}
%


\section{Yang-Baxter equation}\label{sec;YBE}

In this section we briefly summarize some details on the checks of the 
Yang-Baxter equation (YBE) that we have preformed for the bound state S-matrix. 

Let us first focus on subspace I. The block $\mathscr{X}$ governing 
the scattering in this subspace is required to satisfy YBE on its own right.
Thus we need to consider the following scattering sequences,
\begin{align}
|0,1,k_1,\bar{k}_1 \rangle \otimes |0,1,k_2,\bar{k}_2\rangle \otimes
|0,1,k_3,\bar{k}_3\rangle \xrightarrow{\text{YBE}}  |0,1,m_1,\bar{m}_1 \rangle
\otimes |0,1,m_2,\bar{m}_2\rangle \otimes |0,1,m_3,\bar{m}_3 \rangle
\end{align}
which give the explicit form of the YBE in subspace I,
\begin{align}\label{eqn:appYBE}
& \sum_{n =0 }^{k_1+k_2} \mathscr{X}^{k_1,k_2}_n (z_1,z_2)
\mathscr{X}^{n,k_3}_{m_2}(z_1,z_3)  \mathscr{X}^{k_1+ k_2 -n,
k_3+n-m_2}_{m_1-m_2}(z_2,z_3)  =  \el
& \qquad\qquad\sum_{n =0 }^{k_2+k_3} \mathscr{X}^{m_1-n,n}_{m_2} (z_1,z_2)
\mathscr{X}^{k_1,k_2+k_3-n}_{m_1-n}(z_1,z_3)  \mathscr{X}^{k_2,
k_3}_{n}(z_2,z_3) \,.
\end{align}
We did not attempt to prove this identity in full generality,
but we did check it for a large set of different values  
of the parameters $k_i,m_i$ and bound state numbers and found it to be
perfectly satisfied. 

In a similar way, this approach for checking YBE may be extended to include
the subspaces II and III. For example, acting with 
$S_{12}S_{13}S_{23}-S_{23}S_{13}S_{12}$ on the states of the following form,
\begin{align}
|0,0,k_1,l_1 \rangle \otimes |0,0,k_2,l_2\rangle \otimes
|0,1,k_3,\bar{k}_3\rangle,
\end{align}
will result in a plethora of different types of states and the coefficients will
depend on all three scattering blocks $\mathscr{X,\,Y,\,Z}$. 
Due to the large size of these expressions we have not spelled them out
explicitly here. Nevertheless we have explicitly computed, for different values of the parameters,
various matrix elements of the YBE that include states from all three
subspaces that in general may be written as 
\begin{align}
\langle \text{~out-state~} | \text{~YBE~} |\text{~in-state~} \rangle\,. 
\end{align}
We have performed the checks for a wide range of numerical values of
the representation parameters and in each case it proved to be compatible with
the YBE.
%



\begin{thebibliography}{}

\newcommand{\nlin}[1]{\href{http://xxx.lanl.gov/abs/nlin/#1}{\tt nlin/#1}}
\newcommand{\hepth}[1]{\href{http://xxx.lanl.gov/abs/hep-th/#1}{\tt hep-th/#1}}
\newcommand{\arxid}[1]{\href{http://arxiv.org/abs/#1}{\tt arXiv:#1}}
\newcommand{\condmat}[1]{\href{http://arxiv.org/abs/cond-mat/#1}{\tt
cond-mat/#1}}
%


\bibitem{Hubbard}J. Hubbard, \textit{Electron Correlations in Narrow Energy
Bands}, Proc. Roy. Soc. London
A 276 (1963) 238;

\bibitem{Hub1}M. Rasetti, \textit{The Hubbard Model -- Recent Results}, World
Scientific Singapore, (1991).

\bibitem{Hub2}A. Montorsi, \textit{The Hubbard Model}, World Scientific
Singapore, (1992).

\bibitem{Hub3}V. Korepin, F. E\ss ler, \textit{Exactly Solvable Models of
Strongly Correlated Electrons}, World Scientific Singapore, (1994).

\bibitem{Hub4}F. E\ss ler, H. Frahm, F. Goehmann, A. Klumper and V. Korepin,
\textit{The One-Dimensional Hubbard Model}, Cambridge University Press, (2005).

\bibitem{tJ}J. Spalek, \textit{t-J Model Then and Now: A Personal Perspective
From the Pioneering Times}, 
Acta Physica Polonica A {111}, 409-24 (2007), [\arxid{0706.4236}].

\bibitem{AB}F. C. Alcaraz and R. Z. Bariev, \textit{Interpolation Between
Hubbard and Supersymmetric t-J models: Two-parameter Integrable Models of
Correlated Electrons}, J. Phys. A32, L483 (1999), [\condmat{9908265}].

\bibitem{AdSReview}N. Beisert et al, \textit{Review of AdS/CFT Integrability: An
Overview}, [\arxid{1012.3982}].

\bibitem{BAnalytic}N. Beisert, \textit{The Analytic Bethe Ansatz for a Chain
with Centrally Extended su$(2|2)$ symmetry}, 
J.\ Stat.\ Mech.\  {0701} (2007) P01017, [\nlin{0610017}].

\bibitem{BK}N. Beisert, P. Koroteev, \textit{Quantum Deformations of the
One-Dimensional Hubbard Model}, 
J.Phys.A41:255204, 2008, [\arxid{0802.0777}].

\bibitem{MartinsMelo}  M.~J.~Martins, C.~S.~Melo, \textit{The Bethe ansatz
approach for factorizable centrally extended S-matrices}, Nucl.\ Phys.\  { B785
} (2007)  246-262 [\arxid{0703086}].

\bibitem{Shastry}B. S. Shastry, \textit{Exact Integrability of the
One-Dimensional Hubbard Model},
Phys. Rev. Lett. 56, 2453 (1986).

\bibitem{UK}D.~B.~Uglov and V.~E.~Korepin, \textit{The Yangian symmetry of the
Hubbard model},
Phys.\ Lett.\  A {190} (1994) 238, [\arxid{hep-th/9310158}].

\bibitem{BeisertYangian}N.~Beisert, \textit{The S-Matrix of AdS / CFT and
Yangian symmetry}, PoS { SOLVAY } (2006) 002. [\arxid{0704.0400} ].

\bibitem{BeisertFundamental} N.~Beisert, \textit{The SU$(2|2)$ Dynamic
S-Matrix},  Adv.\ Theor.\ Math.\ Phys.\  { 12 } (2008)  945.  [\arxid{0511082}].

\bibitem{AFZ} G.~Arutyunov, S.~Frolov, M.~Zamaklar, \textit{The
Zamolodchikov-Faddeev Algebra for $\ads$ Superstring}, JHEP { 0704 } (2007) 
002.  [\arxid{0612229}].

\bibitem{BJ} Z.~Bajnok, R.~A.~Janik, \textit{Four-loop Perturbative Konishi from
Strings and Finite Size Effects For Multiparticle States},  Nucl.\ Phys.\  {
B807 } (2009)  625-650. [\arxid{0807.0399}].

\bibitem{Janik}R. Janik, \textit{The $AdS_5 \times S^5$
Superstring Worldsheet S-Matrix and Crossing Symmetry}, Phys. Rev. D73:086006,
(2006) [\hepth{0603038}].

\bibitem{AFS} G.~Arutyunov, S.~Frolov, M.~Staudacher, \textit{Bethe Ansatz For
Quantum Strings}, JHEP {0410 } (2004) 016. [\arxid{0406256}]. 

\bibitem{BES}  N.~Beisert, B.~Eden, M.~Staudacher, \textit{Transcendentality and
Crossing},  J.\ Stat.\ Mech.\  { 0701 } (2007)  P01021. [\arxid{0610251}].

\bibitem{AFBound} G. Arutyunov, S. Frolov, \textit{The S-Matrix of String Bound
States}, 
Nucl.Phys.B804:90-143, 2008, [\arxid{0803.4323}].

\bibitem{MdLrmat}  M.~de Leeuw, \textit{Bound States, Yangian Symmetry and
Classical r-matrix for the $\ads$ Superstring}, JHEP { 0806 } (2008)  085.
[\arxid{0804.1047}].

\bibitem{Bsu22}N. Beisert,\textit{ The Analytic Bethe Ansatz
for a Chain with Centrally Extended $su(2|2)$ Symmetry}, J.Stat.Mech.0701:P017,
(2007) [\nlin{SI}{0610017}].

\bibitem{Dorey} N.~Dorey, \textit{Magnon Bound States and the AdS/CFT
Correspondence}, J.\ Phys.\ A { A39 } (2006)  13119-13128. [\arxid{0604175}].

\bibitem{CDO} H.~-Y.~Chen, N.~Dorey, K.~Okamura, \textit{Dyonic Giant Magnons},
JHEP { 0609 } (2006)  024. [\arxid{0605155}].

\bibitem{CDO2} H.~-Y.~Chen, N.~Dorey, K.~Okamura, \textit{On the Scattering of
Magnon Boundstates}, JHEP {0611 } (2006)  035. [\arxid{0608047}].

\bibitem{ALT}G. Arutyunov, M. de Leeuw, A. Torrielli, \textit{The Bound State
S-Matrix for $\ads$ Superstring}, 
Nucl.Phys.B819:319-350, 2009, [\arxid{0902.0183}].

\bibitem{MM}T. Matsumoto, S. Moriyama, \textit{An Exceptional
Algebraic Origin of the AdS/CFT Yangian Symmetry}, JHEP 0804 (2008)
022, [\arxid{0803.1212}]. 

\bibitem{ST}F. Spill, A. Torrielli, \textit{On Drinfeld's Second
Realization of the AdS/CFT su(2$\vert$2) Yangian}, J.Geom.Phys.59:489-502
(2009), [\arxid{0803.3194}].

\bibitem{ALTlong}
G. Arutyunov, M. de Leeuw, A. Torrielli, \textit{On Yangian and Long
Representations of the Centrally Extended $\alg{su}(2|2)$ Superalgebra}, JHEP,
1006, 033, 2010, [\arxid{0912.0209}].

\bibitem{ALTblocks}, G.~Arutyunov, M.~de Leeuw and A.~Torrielli,
\textit{Universal Blocks of the AdS/CFT Scattering Matrix}, 
JHEP {0905} (2009) 086, [\arxid{0903.1833}].

\bibitem{MMT} T.~Matsumoto, S.~Moriyama, A.~Torrielli, \textit{A Secret Symmetry
of the AdS/CFT S-Matrix}, JHEP { 0709 } (2007)  099.  [\arxid{0708.1285}].

\bibitem{DFFR}J. M. Drummond, G. Feverati, L. Frappat, E. Ragoucy,
\textit{Super-Hubbard Models and Applications}, 
JHEP 0705 (2007) 008, [\hepth{0703078}].

\bibitem{Bcl}
N.~Beisert, \textit{The Classical Trigonometric r-Matrix for the
Quantum-Deformed Hubbard
Chain}, J.\ Phys.\ A  { 44} (2011) 265202, [\arxid{1002.1097}].

\bibitem{BGM}N. Beisert, W. Galleas, T. Matsumoto \textit{A Quantum Affine
Algebra for the Deformed Hubbard Chain},
[\arxid{1102.5700}].

\bibitem{Macfarlane} A. J. Macfarlane, \textit{On q Analogs of the Quantum
Harmonic Oscillator and the Quantum Group SU(2)-q}, J.Phys.A, A22:4581, 1989.

\bibitem{Biedenharn} L. C. Biedenharn, \textit{The Quantum Group SU(2)-q and a
q-Analog of the Boson Operators}, J.Phys.A, A22:L873, 1989.

\bibitem{Hayashi} T. Hayashi, \textit{Q Analogs of Clifford and Weyl Algebras:
Spinor and Oscillator Reprsentations of Quantum Enveloping Algebras},
Commun.Math.Phys.127:129-144, 1990.

\bibitem{Chaichian} M. Chaichian, R. Kulish \textit{Quantum Lie Superalgebras
and q-Oscillators}, Phys.Lett., B234:72, 1990.

\bibitem{AFstring}  G.~Arutyunov, S.~Frolov, \textit{String Hypothesis for the
$\ads$ Mirror}, JHEP { 0903 } (2009) 152. [\arxid{0901.1417}]. 

\bibitem{GKV} N.~Gromov, V.~Kazakov, P.~Vieira, \textit{Exact Spectrum of
Anomalous Dimensions of Planar N=4 Supersymmetric Yang-Mills Theory}, Phys.\
Rev.\ Lett.\  { 103 } (2009)  131601.
  [\arxid{0901.3753}].

\bibitem{BFT} D.~Bombardelli, D.~Fioravanti, R.~Tateo, \textit{Thermodynamic
Bethe Ansatz for Planar AdS/CFT: A Proposal},  J.\ Phys.\ A {A42} (2009) 
375401. [\arxid{0902.3930}].

\bibitem{AFTBA}  G.~Arutyunov, S.~Frolov, \textit{Thermodynamic Bethe Ansatz for
the $\ads$ Mirror Model}, JHEP { 0905 } (2009)  068 [\arxid{0903.0141}].

\bibitem{HT} B.~Hoare, A.~A.~Tseytlin, \textit{Towards the Quantum S-matrix of
the Pohlmeyer Reduced Version of $\ads$ Superstring Theory}, Nucl.\ Phys.\ 
{B851 } (2011)  161-237, [\arxid{1104.2423}].

\bibitem{HHM}  B.~Hoare, T.~J.~Hollowood, J.~L.~Miramontes, \textit{A
Relativistic Relative of the Magnon S-Matrix}, [\arxid{1107.0628}].

\bibitem{Drinfeld}  V.~G.~Drinfeld, \textit{Quasi Hopf Algebras},  Alg.\ Anal.\ 
{1N6 } (1989)  114-148.

\bibitem{Petersen}
J-U. Petersen, \textit{Representations at a Root of Unity of q-Oscillators and
Quantum Kac-Moody Algebras}, 1994, Ph.D Thesis, [\arxid{hep-th/9409079}].

\bibitem{ChaichianBook} M. Chaichian,  A.P. Demichev, \textit{Introduction to
Quantum Groups}, 1996.

\bibitem{KR}A.~N.~Kirillov, N.~Y.~.Reshetikhin, \textit{Representations of the
Algebra $U(q)(sl(2))$ q-Orthogonal Polynomials and Invariants of Links},
  Kohno, T. (ed.): New developments in the theory of knots, 202-256 (1991).

\bibitem{MN}R. Murgan and R. I. Nepomechie, \textit{q-Deformed $su(2|2)$
Boundary S-Matrices via the ZF
Algebra}, JHEP 0806 (2008) 096, [\arxid{0805.3142}].

\bibitem{AN}C. Ahn, R. I. Nepomechie, \textit{Yangian Symmetry
and Bound-States in AdS/CFT Boundary Scattering}, JHEP 1005 (2010)
016, [\arxid{1003.3361}].

\bibitem{MR1}N. MacKay, V. Regelskis, \textit{Yangian Symmetry
of the Y=0 Maximal Giant Graviton}, JHEP 1012 (2010) 076, [\arxid{1010.3761}].

\bibitem{MR2}N. MacKay, V. Regelskis, \textit{Reflection Algebra, Yangian
Symmetry and Bound-States in AdS/CFT}, [\arxid{1101.6062}].

\bibitem{LeeuwThesis}  M.~de Leeuw, \textit{The S-matrix of the $AdS_5 x S^5$
superstring}, [\arxid{1007.4931}].


\end{thebibliography}
\end{document}